\documentclass[10pt,journal,compsoc]{IEEEtran}

\ifCLASSOPTIONcompsoc
\usepackage[caption=false,font=footnotesize,labelfont=sf,textfont=sf]{subfig}
\else
\usepackage[caption=false,font=footnotesize]{subfig}
\fi

\usepackage[pdftex]{graphicx}
\usepackage{epstopdf}
\DeclareGraphicsExtensions{.eps}

\usepackage{amsmath}
\usepackage{amssymb}
\usepackage{amsthm}
\usepackage{url}
\usepackage{listings}
\usepackage{color}
\usepackage{algorithm}
\usepackage{algpseudocode}
\usepackage{verbatim}
\usepackage{tikz}
\usepackage{tikz-qtree}
\usepackage{wasysym}
\usepackage{mathtools}

\algnewcommand\algorithmicinput{\textbf{Input:}}
\algnewcommand\INPUT{\item[\algorithmicinput]}
\algnewcommand\algorithmicoutput{\textbf{Output:}}
\algnewcommand\OUTPUT{\item[\algorithmicoutput]}
\algnewcommand{\IIf}[1]{\State\algorithmicif\ #1\ \algorithmicthen}
\algnewcommand{\EndIIf}{\unskip\ \algorithmicend\ \algorithmicif}

\newtheorem{theorem}{Theorem}[section]
\newtheorem{lemma}[theorem]{Lemma}
\newtheorem{proposition}[theorem]{Proposition}

\newtheorem*{proposition*}{Proposition}

\definecolor{darkcyan}{rgb}{0.0, 0.55, 0.55}

\begin{document}

\title{Bi-objective Optimization of Data-parallel Applications on Heterogeneous Platforms for Performance and Energy via Workload Distribution}
\author{Hamidreza~Khaleghzadeh,
		Muhammad~Fahad,
		Arsalan~Shahid,
        Ravi~Reddy~Manumachu,
        and~Alexey~Lastovetsky
\IEEEcompsocitemizethanks{\IEEEcompsocthanksitem H. Khaleghzadeh, M. Fahad, A.Shahid, R. Reddy and A. Lastovetsky are with the School of Computer Science, University College Dublin, Belfield, Dublin 4, Ireland.\protect\\
E-mail: hamidreza.khaleghzadeh@ucdconnect.ie, muhammad.fahad@ucdconnect.ie, arsalan.shahid@ucdconnect.ie, ravi.manumachu@ucd.ie, alexey.lastovetsky@ucd.ie}
\thanks{}}



\IEEEtitleabstractindextext{%
\begin{abstract}
Performance and energy are the two most important objectives for optimization on modern parallel platforms. Latest research demonstrated the importance of workload distribution as a key decision variable in the bi-objective optimization of data-parallel applications for performance and energy on homogeneous multicore CPU clusters. We show in this work that moving from single objective optimization for performance or energy to their bi-objective optimization on heterogeneous processors results in a tremendous increase in the number of optimal solutions (workload distributions) even for the simple case of linear performance and energy profiles. We then study full performance and energy profiles of two real-life data-parallel applications and find that they exhibit shapes that are non-linear and complex enough to prevent good approximation of them as analytical functions for input to exact algorithms or optimization softwares for determining the globally Pareto-optimal front.

We, therefore, propose a solution method solving the bi-objective optimization problem on heterogeneous processors and comprising of two principal components. The first component is an efficient and exact global optimization algorithm. The algorithm takes as an input most general discrete performance and dynamic energy functions that accurately and realistically account for resource contention and NUMA inherent in modern parallel platforms. The algorithm is also used as a building block to solve the bi-objective optimization problem for performance and total energy. The second component is a novel methodology employed to build the discrete dynamic energy profiles of individual computing devices, which are input to the algorithm. The methodology is based purely on system-level measurements and addresses a fundamental challenge, which is to accurately model the energy consumption by a hybrid scientific data-parallel application executing on a heterogeneous HPC platform containing different computing devices such as CPU, GPU, and Xeon PHI.

We experimentally analyse the proposed solution method using two data-parallel applications, matrix multiplication and 2D fast Fourier transform (2D-FFT), and show that our solution method determines a superior Pareto-optimal front containing all the load imbalanced solutions that are totally ignored by \emph{load balancing} methods and best load balanced solutions.
\end{abstract}

\begin{IEEEkeywords}
heterogeneous platforms, data-parallel applications, data partitioning, performance optimization, energy optimization, bi-objective optimization, workload distribution, multicore CPU, GPU, Intel Xeon Phi
\end{IEEEkeywords}}

\maketitle

\IEEEpeerreviewmaketitle

\IEEEraisesectionheading{\section{Introduction}\label{sec:introduction}}

Performance and energy are the two most important objectives for optimization on modern parallel platforms such as supercomputers, heterogeneous HPC clusters, and cloud computing infrastructures (\cite{Fard2012,Kessaci2013,Durillo2014,rossi2017eco}).

State-of-the-art solutions for bi-objective optimization problem for performance and energy on heterogeneous HPC platforms can be broadly classified into \emph{system-level} and \emph{application-level} categories. The objectives used in these solutions are performance and total energy. Briefly, the total energy consumption is the sum of dynamic and static energy consumptions. We define the static energy consumption as the energy consumed by the platform without the application execution. Dynamic energy consumption is calculated by subtracting this static energy consumption from the total energy consumed by the platform during the application execution. 

System-level solution methods aim to optimize performance and energy of the environment where the applications are executed. The methods employ application-agnostic models and hardware parameters as decision variables. The dominant decision variable in this category is Dynamic Voltage and Frequency Scaling (DVFS). Majority of the works in this category can be further grouped as follows: a). Methods optimizing for performance under a power cap constraint (or energy budget) or optimizing for energy under an execution time constraint \cite{Yu2015,gholkar2016power,Rountree2017}. They determine a partial Pareto-optimal front of solutions by applying the power cap or an execution time constraint and then select the best configuration fulfilling an user-specific criterion. b). Methods solving unconstrained bi-objective optimization for performance and energy \cite{Kessaci2013,Durillo2014,Kolodziej2015}. They build the full globally Pareto-optimal front of solutions.

Application-level solution methods proposed in \cite{Subramaniam2010,Demmel2013,Lang2014,chakrabarti2017pareto,LastovetskyReddy2017,manumachu2018bi,manumachu2018bicpe} use application-level parameters as decision variables and application-level models for predicting the performance and energy consumption of applications to solve the bi-objective optimization problem. The application-level parameters include the number of threads, number of processors, loop tile size, workload distribution, etc. The methods in \cite{Subramaniam2010,Demmel2013,Lang2014} do not consider workload distribution as a decision variable. The methods proposed in \cite{LastovetskyReddy2017,manumachu2018bi,manumachu2018bicpe} demonstrate by executing real-life data-parallel applications on modern multicore CPUs that the functional relationships between performance and workload size and between energy and workload size have complex (non-linear) properties and show that workload distribution has become an important decision variable that can no longer be ignored. The methods target homogeneous HPC platforms. \cite{chakrabarti2017pareto} consider the effect of heterogeneous workload distribution on bi-objective optimization of data analytics applications by simulating heterogeneity on homogeneous clusters. The performance is represented by a linear function of problem size and the total energy is predicted using historical data tables.

In this work, we study two bi-objective optimization problems for data-parallel applications on \emph{heterogeneous} HPC systems. The problems aim to optimize the parallel execution of a given workload $n$ by a set of $p$ heterogeneous processors. The first optimization problem, \emph{HEPOPT}, has two objectives, execution time and dynamic energy, and one decision variable, the \emph{workload distribution}. The second optimization problem, \emph{HTPOPT}, has the same decision variable, the \emph{workload distribution}, but objectives, which are performance and total energy.

The motivation for the study comes from our observation of the effect of heterogeneity on the solution space as we move from single objective optimization for performance or energy to bi-objective optimization for performance and energy for the simple case where the execution time and dynamic energy functions are linear.

\begin{figure}[!t]
	\centering
	\includegraphics[width=3.5in]{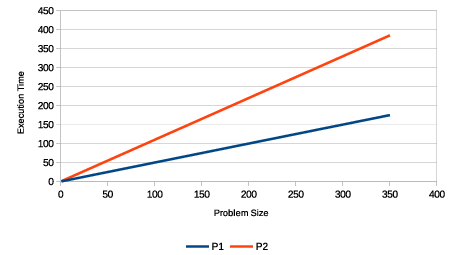}
	\caption{Linear execution time functions of the processors $P_1$ and $P_2$.}
	\label{fig:linearetime}
\end{figure}

\begin{figure}[!t]
	\centering
	\includegraphics[width=3.5in]{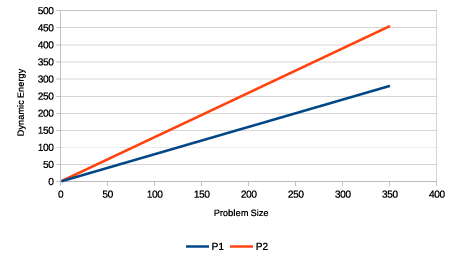}
	\caption{Linear dynamic energy functions of the processors $P_1$ and $P_2$.}
	\label{fig:linearenergy}
\end{figure}

Consider two processors $P_1$ and $P_2$, whose linear execution time and dynamic energy functions are shown in the Figures \ref{fig:linearetime} and \ref{fig:linearenergy}. The functions are real-life profiles of a data-parallel matrix multiplication application executed using a single core of a multicore CPU. For a given input workload size $n$, an exact algorithm determines the globally Pareto-optimal front of solutions (distributions $(x_1,x_2)$ of workload $n$ where $x_1+x_2=n$). The solution for single objective optimization for performance (minimizing the execution time of computations during the parallel execution of the workload) is the load balanced solution where all the processors involved in the parallel execution of a given workload have equal execution times. The solution for single objective optimization for dynamic energy (minimizing the total dynamic energy) allocates the entire workload to the most energy efficient processor, $P_1$.

\begin{figure}[!t]
	\centering
	\includegraphics[width=3.5in]{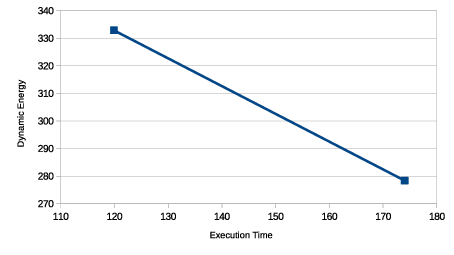}
	\caption{Globally Pareto-optimal front of solutions for a workload size $348$. The front is linear. The end points are the solutions for single objective optimization for performance and for dynamic energy.}
	\label{fig:linear_pareto_max}
\end{figure}

Now consider solving \emph{HEPOPT} using the two processors. The globally Pareto-optimal front shown in Figure \ref{fig:linear_pareto_max} is linear containing an infinite number of solutions. The endpoints are the solutions for single objective optimization for performance and for dynamic energy. We prove (in Section \ref{sec:append}) that for an arbitrary number of processors with linear execution time and dynamic energy functions, the globally Pareto-optimal front is linear and contains an infinite number of solutions out of which one solution is load balanced while the rest are load imbalanced.

We thus discover that moving from single objective optimization for performance or dynamic energy to bi-objective optimization for performance and dynamic energy on heterogeneous processors results in a drastic increase in the number of optimal solutions for the simple case of linear performance and energy profiles, with practically all the solutions load imbalanced.

\begin{table}
\caption{HCLServer1: Specifications of the Intel Haswell multicore CPU, Nvidia K40c, and Intel Xeon Phi 3120P.}
\label{table:hclserver1}
\centering
\begin{tabular}{ |l|l| }
  \hline
  \multicolumn{2}{|c|}{\textbf{Intel Haswell E5-2670V3}} \\ \hline
	No. of cores per socket & 12 \\ \hline
	Socket(s) & 2 \\ \hline
	CPU MHz & 1200.402 \\ \hline
    L1d cache, L1i cache  & 32 KB, 32 KB \\ \hline
    L2 cache, L3 cache & 256 KB, 30720 KB \\ \hline
    Total main memory & 64 GB DDR4 \\ \hline
	Memory bandwidth & 68 GB/sec \\ \hline
  \multicolumn{2}{|c|}{\textbf{NVIDIA K40c}} \\ \hline
	No. of processor cores & 2880 \\ \hline
	Total board memory & 12 GB GDDR5 \\ \hline
	L2 cache size & 1536 KB \\ \hline
	Memory bandwidth & 288 GB/sec \\ \hline
  \multicolumn{2}{|c|}{\textbf{Intel Xeon Phi 3120P}} \\ \hline
	No. of processor cores & 57 \\ \hline
	Total main memory & 6 GB GDDR5 \\ \hline
	Memory bandwidth & 240 GB/sec \\ \hline
\end{tabular}
\end{table}

\begin{table}
\caption{HCLServer2: Specifications of the Intel Skylake multicore CPU and Nvidia P100 PCIe.}
\label{table:hclserver2}
\centering
\begin{tabular}{ |l|l| }
  \hline
  \multicolumn{2}{|c|}{\textbf{Intel Xeon Gold 6152}} \\ \hline
	Socket(s) & 1 \\ \hline
	Cores per socket & 22 \\ \hline
       L1d cache, L1i cache  & 32 KB, 32 KB \\ \hline
       L2 cache, L3 cache & 256 KB, 30976 KB \\ \hline        
	Main memory &  96 GB \\ \hline
  \multicolumn{2}{|c|}{\textbf{NVIDIA P100 PCIe}} \\ \hline
	No. of processor cores & 3584 \\ \hline
	Total board memory & 12 GB CoWoS HBM2 \\ \hline
	Memory bandwidth & 549 GB/sec \\ \hline
\end{tabular}
\end{table}

Motivated by this finding, we study the performance and dynamic energy profiles of two data-parallel applications executed on two connected heterogeneous multi-accelerator NUMA nodes, \emph{HCLServer01} and \emph{HCLServer02}. We observe that the shapes of the speed and dynamic energy functions are non-linear and complex, and therefore difficult to approximate as analytical functions that can be used as inputs to exact mathematical algorithms or optimization softwares for determining the globally Pareto-optimal front. We, therefore, propose an efficient global optimization algorithm \emph{HEPOPTA} that takes as input, most general discrete execution time and dynamic energy functions to determine the globally Pareto-optimal front. Using this algorithm, we present now an analysis of the quality of the fronts for the two applications. 

The first node, \emph{HCLServer01}, consists of an Intel Haswell multicore CPU involving 24 physical cores with 64 GB main memory, which is integrated with two accelerators, one Nvidia K40c GPU and one Intel Xeon Phi 3120P, whose specifications are shown in Table \ref{table:hclserver1}. \emph{HCLServer02} contains an Intel Skylake multicore CPU consisting of 22 cores and 96 GB main memory. The multicore CPU is integrated with one Nvidia P100 GPU, whose specifications can be found in Table \ref{table:hclserver2}. Each accelerator is connected to a dedicated host core via a separate PCI-E link.

A data-parallel application executing on this heterogeneous hybrid platform, consists of a number of kernels (generally speaking, multithreaded), running in parallel on different computing devices of the platform. The proposed algorithm for solving \emph{HEPOPT} requires individual performance and energy profiles of all the kernels. Due to tight integration and severe resource contention in heterogeneous hybrid platforms, the load of one computational kernel in a given hybrid application may significantly impact the performance of others to the extent of preventing the ability to model the performance and energy consumption of each kernel in hybrid applications individually \cite{Zhong2015}. To address this issue, we restrict our study in this work to such configurations of hybrid applications, where individual kernels are coupled loosely enough to allow us to build their individual performance and energy profiles with the accuracy sufficient for successful application of the proposed algorithms. To achieve this, we only consider configurations where no more than one CPU kernel or accelerator kernel is running on the corresponding device. In order to apply our optimization algorithms, each group of cores executing an individual kernel of the application is modelled as an abstract processor \cite{Zhong2015} so that the executing platform is represented as a set of heterogeneous abstract processors. We make sure that the sharing of system resources is maximized within groups of computational cores representing the abstract processors and minimized between the groups. This way, the contention and mutual dependence between abstract processors are minimized.

We thus model \emph{HCLServer01} by three abstract processors, CPU\_1, GPU\_1 and PHI\_1. CPU\_1 represents 22 (out of total 24) CPU cores. GPU\_1 involves the Nvidia K40c GPU and a host CPU core connected to this GPU via a dedicated PCI-E link. PHI\_1 is made up of one Intel Xeon Phi 3120P and its host CPU core connected via a dedicated PCI-E link. In the same manner, \emph{HCLServer02} is modelled by two abstract processors, CPU\_2 and GPU\_2. Since there should be a one-to-one mapping between the abstract processors and computational kernels, any hybrid application executing on the servers in parallel should consist of five kernels, one kernel per computational device.

Because the abstract processors contain CPU cores that share some resources such as main memory and QPI, they cannot be considered completely independent. Therefore, the performance of these loosely-coupled abstract processors must be measured simultaneously, thereby taking into account the influence of resource contention \cite{Zhong2015} .

To model the performance of a parallel application and build its speed functions, the execution time of any computational kernel can be measured accurately using high precision processor clocks. There is however no such effective equivalent for measuring the energy consumption. There are two dominant approaches to determine energy consumption: a). Hardware-based, such as using on-chip power sensors or physical measurements using external power meters, and b). Software-based, such as energy predictive models using performance monitoring counters (PMCs). While energy predictive models provide the decomposition of energy consumption at the component level, they exhibit poor prediction accuracy and demonstrate high implementation complexity (\cite{McCullough2011, Ken2017, Arsalan2017}). Physical measurements using power meters are accurate but they do not provide a fine-grained decomposition of the energy consumption during the application run in a hybrid platform. We propose a novel methodology to determine this decomposition, which employs only system-level energy measurements using power meters. The methodology allows us to build discrete dynamic energy functions of abstract processors with sufficient accuracy for the application of the proposed optimization algorithms in our use cases. In our future work, we plan to use measurements provided by on-chip power sensors to improve the accuracy of the methodology and to reduce the execution time to build the energy profiles.

In our first use case, we experiment with a matrix multiplication application, DGEMM. The application computes $C = \alpha \times A \times B + \beta \times C$, where $A$, $B$, and $C$ are matrices of size $m \times n$, $n \times n$, and $m \times n$, and $\alpha$ and $\beta$ are constant floating-point numbers. The application uses Intel MKL DGEMM for CPUs, ZZGEMMOOC out-of-card package \cite{khaleghzadeh2018out} for Nvidia GPUs and XeonPhiOOC out-of-card package \cite{khaleghzadeh2018out} for Intel Xeon Phis. ZZGEMMOOC and XeonPhiOOC packages reuse CUBLAS and MKL BLAS for in-card DGEMM calls. The out-of-card packages allow the GPUs and Xeon Phis to execute computations of arbitrary size. The Intel MKL and CUDA versions used on \emph{HCLServer01} are 2017.0.2 and 7.5, and on \emph{HCLServer02} are 2017.0.2 and 9.2.148. Workload sizes range from $64\times10112$ to $28800 \times10112$ with a step size of $64$ for the first dimension $m$. The speed of execution of a given problem size $m \times n$ is calculated as $(2 \times m \times n^2)/t$ where $t$ is the execution time.

\begin{figure}[!t]
	\centering
	\includegraphics[width=3.5in]{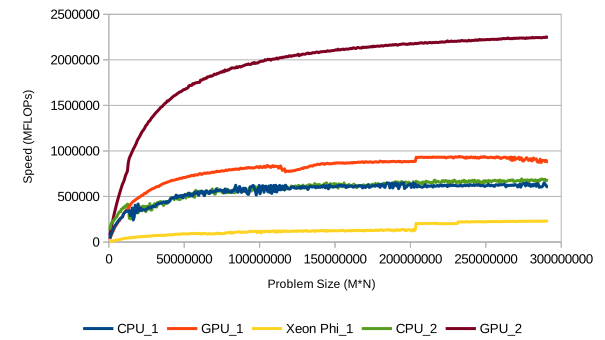}
	\caption{Speed functions of heterogeneous DGEMM application executing on \emph{HCLServer01} and \emph{HCLServer02}. Each data point shows the speed for the execution of a problem size $M \times N$, where $M$ ranges from $64$ to $28800$ and $N$ is $10112$.}
	\label{fig:dgemmfullspeed}
\end{figure}

\begin{figure}[!t]
	\centering
	\includegraphics[width=3.5in]{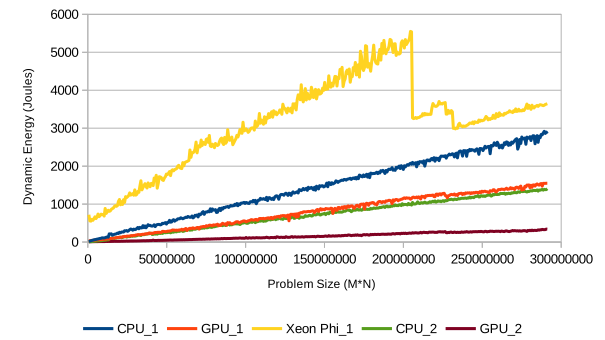}
	\caption{Dynamic energy functions of heterogeneous DGEMM application executing on \emph{HCLServer01} and \emph{HCLServer02} for problem sizes of $M \times N$, where $M$ ranges from $64$ to $28800$ and $N$ is $10112$.}
	\label{fig:dgemmfullenergy}
\end{figure}

Figures \ref{fig:dgemmfullspeed} and \ref{fig:dgemmfullenergy} show the speed and dynamic energy functions of CPU\_1, GPU\_1, Phi\_1 abstract processors, on \emph{HCLServer01}, and CPU\_2 and GPU\_2 abstract processors, on \emph{HCLServer02}. For each data point in these functions, the experiments are repeated until sample means of all the five kernels running on the abstract processors fall in the confidence interval of 95\%. Our experimental methodology is detailed in Section \ref{sec:append}. The shapes of the discrete speed and energy functions are smooth.

\begin{figure}[!t]
	\centering
	\includegraphics[width=3.5in]{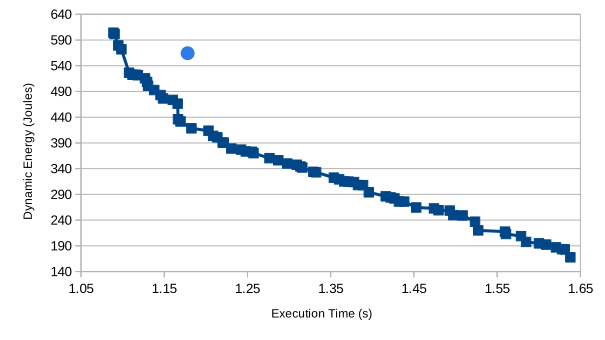}
	\caption{Pareto-front solutions of heterogeneous DGEMM application for a given workload size $w=17152\times10112$. Blue circle represents the load-balanced solution.}
	\label{fig:dgemm_pareto_17152_10112}
\end{figure} 

Figure \ref{fig:dgemm_pareto_17152_10112} shows the globally optimal Pareto front containing $68$ solutions for the given workload size $w=17152\times10112$. The solutions are the workload distributions employing one or more of the available abstract processors. The workload distribution with the maximum performance has an execution time of $1.08$ seconds and dynamic energy consumption of $604$ joules. The workload distribution with the minimum dynamic energy consumption of $167$ joules has the execution time of $1.63$ seconds. Optimizing for dynamic energy consumption degrades performance by $51\%$ whereas optimizing for execution time increases dynamic energy consumption by $260\%$. The load balanced solution is shown by a blue circle in the figure.

In our second use case, we study the performance and dynamic energy profiles of a 2D fast Fourier transform (2D-FFT) application on \emph{HCLServer01} and \emph{HCLServer02}. The application computes 2D-DFT of a complex signal matrix of size $m \times n$. It employs Intel MKL FFT routines for CPUs and Xeon Phis, and CUFFT routines for Nvidia GPUs. All computations are in-card. Workloads range from $1024\times51200$ to $10000\times51200$ with the step size 16 for $m$. The experimental data set does not include problem sizes that cannot be factored into primes less than or equal to 127. For these problem sizes, CUFFT for GPU gives failures. The speed of execution of a 2D-DFT of size $m \times n$ is calculated as $(2.5 \times m \times n \times \log_2 (m \times n))/t$ where $t$ is the execution time. 

\begin{figure}[!t]
	\centering
	\includegraphics[width=3.5in]{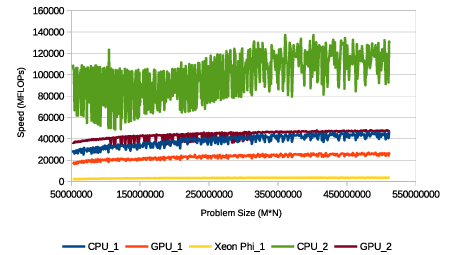}
	\caption{Speed functions of heterogeneous 2D-FFT application executing on \emph{HCLServer01} and \emph{HCLServer02}. The application computes the fast Fourier transpose of a matrix of size $M \times N$, where $M$ ranges from $1024$ to $10000$ and $N$ is $51200$.}
	\label{fig:fftfullspeed}
\end{figure} 

\begin{figure}[!t]
	\centering
	\includegraphics[width=3.5in]{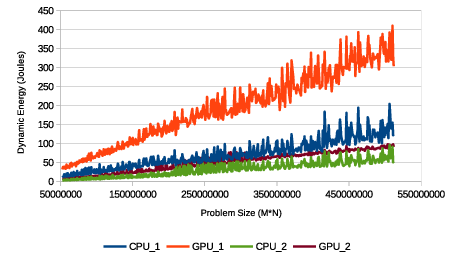}
	\caption{Dynamic energy functions of heterogeneous 2D-FFT application executing on \emph{HCLServer01} and \emph{HCLServer02}. In this figure, the dynamic energy profile for Phi\_1 is ignored since it consumes 10 times more energy and dominates the other profiles. The application calculates the FFT of a matrix size $M \times N$, where $M$ ranges from $1024$ to $10000$ and $N$ is $51200$.}
	\label{fig:fftenergy_wPhi}
\end{figure} 

\begin{figure}[!t]
	\centering
	\includegraphics[width=3.5in]{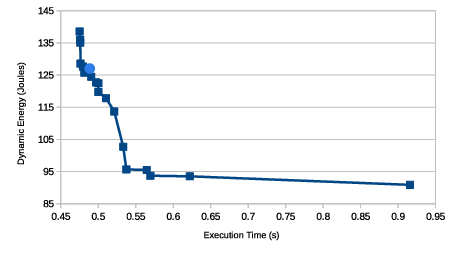}
	\caption{Pareto-front solutions of 2D-FFT for a given workload size, $w=14960\times51200$. Blue circle represents the load-balanced solution.}
	\label{fig:fft_pareto_14960_51200}
\end{figure}

Figures \ref{fig:fftfullspeed} and \ref{fig:fftenergy_wPhi} show the speed and dynamic energy functions of the abstract processors. Here, speed and energy profiles show drastic variations. Figure \ref{fig:fft_pareto_14960_51200} shows the globally optimal Pareto front containing $18$ solutions for the input workload size $w=14960\times51200$. The workload distribution maximizing the performance has the execution time of $0.48$ seconds and dynamic energy consumption of $138$ joules. The workload distribution with the minimal dynamic energy consumption of $90$ joules has the execution time of $0.92$ seconds. Optimizing for dynamic energy consumption alone degrades performance by $92\%$, and optimizing for performance alone increases dynamic energy consumption by $53\%$. The blue circle in the figure shows the load balanced solution, which is close to the Pareto-optimal front of solutions.

We thus observe a good number of trade-off solutions for performance and dynamic energy when workload distribution is used as the decision variable. We notice however that the number of solutions in the Pareto-optimal front depends on the shapes (and smoothness) of the profiles. It is large for smooth profiles compared to non-smooth profiles with severe variations. In our future work, we will study theoretically the constraints for non-linear shapes for performance and energy functions when one can expect good trade-off solutions.

The algorithm \emph{HEPOPTA} solving \emph{HEPOPT} takes as inputs, the workload size, $n$; the number of available heterogeneous processors, $p$; $p$ discrete performance functions (one for each processor); and $p$ discrete dynamic energy functions (one for each processor). The performance and energy functions are functions of workload size. The algorithm returns the globally Pareto-optimal set of solutions for performance and dynamic energy. Each solution in the set is a distribution of workload $n$ between the $p$ heterogeneous processors, which, generally speaking, is not balanced. To the best of the authors' knowledge, none of the traditional approaches to optimization for performance and energy consider non-balanced solutions as optimal. \emph{HEPOPTA} returns globally Pareto-optimal solutions for performance and dynamic energy. We prove its computational complexity of $O(m^3 \times p^3 \times \log_2(m \times p))$, where $m$ represents the cardinality of the discrete functions representing the execution time and dynamic energy functions.

We analyse experimentally this algorithm using the two data-parallel applications, matrix multiplication and 2D-FFT. The average and maximum percentage reductions in execution time and dynamic energy consumption against load balanced workload distribution are ($26\%$,$102\%$) and ($130\%$,$257\%$) for matrix multiplication, and ($7\%$,$44\%$) and ($44\%$,$105\%$) for 2D-FFT. The average and maximum number of globally Pareto-optimal solutions for the two applications are ($55$,$96$) and ($11$,$33$). 

The Pareto-optimal sets contain the best load balanced solutions (mostly one solution in very few cases) whereas the rest of the solutions are load imbalanced. We distinguish between two types of load imbalanced solutions, \emph{strong} and \emph{weak}. We will define the two types in terms of both the workload distribution and the ratio of execution times of load balanced and load imbalanced solutions (called the load imbalance ratio (LIR)). A \emph{strong} load imbalanced solution is one where one or more processors can be assigned zero workloads, the same as for the load balanced solution, but the rest of the processors, different workloads. A \emph{weak} load imbalanced solution represents the case where all the processors are assigned workloads that are different compared to the workloads in the load balanced solution. The LIR for a strong load imbalanced solution is higher than that for a weak load imbalanced solution.  A very high percentage of solutions determined by \emph{HEPOPTA} for the two applications are \emph{strong} load imbalanced where PHI\_1 is given workload of size zero. Therefore, \emph{HEPOPTA} determines a superior Pareto-optimal front containing all \emph{strong} load imbalanced solutions that are totally ignored by \emph{load balancing} approaches (Figures \ref{fig:dgemm_pareto_17152_10112}, \ref{fig:fft_pareto_14960_51200}).

Apart from dynamic energy consumption, the enormous total energy consumption in data centres and big clusters is also a critical constraint. The amount of base energy consumed by idle computers in clusters and clouds is non-negligible \cite{piraghaj2017survey}. To save total energy consumption, the idle computers in clusters, clouds, web-servers, and big data centers are switched off or put in sleep mode \cite{chen2015improving,rossi2017eco,benoit2018reducing}. Therefore, the bi-objective optimization problem for performance and total energy (\emph{HTPOPT}) is important in this context. 

We propose an algorithm \emph{HTPOPTA} solving \emph{HTPOPT}. The inputs to this algorithm are the same as those for \emph{HEPOPTA} and the base power of the platform. It reuses \emph{HEPOPTA} to determine the globally Pareto-optimal solutions for performance and total energy. Its time complexity is the same as \emph{HEPOPTA}. We demonstrate that minimisation of the dynamic energy consumption may not necessarily minimise the total energy consumption. The average and maximum difference in total energy consumption between the dynamic-energy optimal and total-energy optimal solutions is ($11\%$,$37\%$) for matrix multiplication, and ($29\%$,$106\%$) for 2D-FFT.

The main original contributions of this work are:
\begin{itemize}
	\item We present the first study on bi-objective optimization for performance and energy of data-parallel applications on heterogeneous processors through optimal workload distribution.
	\item We discover that moving from single objective optimization for performance or energy to bi-objective optimization for performance and energy on heterogeneous processors results in a drastic increase in the number of optimal solutions in the case of linear performance and energy profiles, with practically all the solutions load imbalanced. We prove that for an arbitrary number of processors with linear execution time and dynamic energy functions, the globally Pareto-optimal front is linear and contains an infinite number of solutions out of which one solution is load balanced while the rest are load imbalanced.
	\item Model-based data partitioning algorithms \emph{HEPOPTA} and \emph{HTPOPTA} for solving the bi-objective optimization problem for execution time and dynamic energy and execution time and total energy for data-parallel applications on heterogeneous HPC platforms. The algorithms take as input discrete speed and dynamic energy functions with any arbitrary shape. The algorithms return the globally Pareto-optimal set of, generally speaking, non-balanced solutions.
	\item A methodology to determine decomposition of dynamic energy consumption using system-level measurements for heterogeneous hybrid servers (integrating a multicore CPU, a GPU and a Xeon Phi) with sufficient accuracy, and experimental validation of the methodology on two modern heterogeneous hybrid servers.
	\item Experimental study of the applicability of \emph{HEPOPTA} and \emph{HTPOPTA} to optimization of real-life state-of-the-art data-parallel applications on two connected hybrid heterogeneous multi-accelerator servers consisting of multicore CPUs, GPUs, and Intel Xeon Phi. We demonstrate that solutions provided by these algorithms significantly improve the performance and reduce the energy consumption of matrix multiplication and 2D fast Fourier transform in comparison with the load-balanced configuration of the applications.
	\item We demonstrate that the proposed solution methods determine a better Pareto-optimal front containing all the load imbalanced solutions that are totally ignored by \emph{load balancing} approaches. We also show that the globally Pareto-optimal front determined by the solution methods contains all the best load balanced solutions in the sense that any other load balanced solution will be sub-optimal. 
\end{itemize}

The rest of the paper is organized as follows. Related work is discussed in section \ref{sec:related-works}. Section \ref{sec:formulations} contains the formulation of the bi-objective optimization problem for performance and dynamic energy, \emph{HEPOPT}. Section \ref{sec:methodology} presents our algorithm, \emph{HEPOPTA}, solving the problem. In section \ref{sec:total-eng-optimiztion}, we formulate and propose our algorithm \emph{HTPOPTA} solving the bi-objective optimization problem for performance and total energy, \emph{HTPOPT}. In section \ref{sec:Energy Modeling}, our device-level approach for dynamic energy modelling is illustrated. We present the experimental results for \emph{HEPOPTA} and \emph{HTPOPTA} in section \ref{sec:exprimental-results}. Finally, we conclude the paper in section \ref{sec:conclusion}.

\section{Related Work}	\label{sec:related-works}

Realistic and accurate performance and energy models of computations are key building blocks for data partitioning algorithms solving the bi-objective optimization problem for performance and energy. We cover them first in our literature survey. We follow this with few notable methods solving bi-objective optimization problem on HPC platforms.

\subsection{Performance Models of Computation}

Performance models of computations can be classified into analytical and non-analytical categories.

Analytical models use techniques such as linear regression, analysing patterns of computation and memory accesses, and static code analysis to estimate performance for CPUs and accelerators \cite{kim2011performance,shen2016workload}. In the non-analytical category, the most simple model is a constant performance model (CPM) where different notions such as normalized cycle time, normalized processor speed, average execution time, task computation time, etc. characterize the speed of an application \cite{Kalinov2001,Beaumont2001}. In CPMs, no dependence is assumed between the performance of a processor and the workload size.

CPMs are too simplistic to accurately model the performance of data-parallel applications executing on modern heterogeneous platforms. The most advanced load balancing algorithms employ functional performance models (FPMs) that are application-specific and that represent the speed of a processor by a continuous function of problem size \cite{Lastovetsky2004,lastovetsky2005data,Lastovetsky2007}. The FPMs capture realistically and accurately the real-life behaviour of applications executing on nodes consisting of uniprocessors (single-core CPUs).

The complex nodal architecture of modern HPC systems comprising of tightly integrated processors with inherent severe resource contention and NUMA pose serious challenges to load balancing algorithms based on the FPMs. These inherent traits result in significant variations (drops) in the performance profiles of parallel applications executing on these platforms thereby violating the assumptions on the shapes of the performance profiles considered by the FPM-based load balancing algorithms. In \cite{lastovetsky2017model,LastovetskyReddy2017,khaleghzadeh2018novel}, novel model-based data partitioning algorithms are proposed that employ load imbalancing parallel computing method to address the new challenges.

\subsection{Energy Modelling Techniques}

There are two dominant approaches to provide an accurate measurement of energy consumption during an application execution \cite{fahad2019comparative}: a). Physical measurements using external power meters or on-chip power sensors, and b). Energy predictive models. While the first approach is known to be accurate, it can only provide the measurement at a computer level and therefore lacks the ability to provide a fine-grained component-level decomposition of the energy consumption of an application. This decomposition is critical to data partitioning algorithms optimizing the application for energy.

\noindent \textbf{\textit{Energy Predictive Models:}} The existing energy predictive models predominantly use Performance Monitoring Counts (PMCs) to predict energy consumption during application execution. A typical approach is to model the energy consumption of a hardware component (such as CPU, DRAM, fans, disks (HDD) etc.) using linear regression of the performance events occurring in the component during application execution.

\noindent \textit{Energy Predictive Models for CPUs:}
Component level energy predictive models based on high positively correlated performance events such as integer operations, floating-point operations, and cache misses include  \cite{Heath2005}, \cite{Economou2006}, \cite{Bircher2012}. They construct models for different hardware components such as CPU, disk, and network based on their utilization. \cite{Basmadjian2011} construct a power model of a server using the summation of power models of its components: the processor (CPU), memory (RAM), fans, and disk (HDD). \cite{LastovetskyReddy2017} propose a model representing the energy consumption of a multicore CPU by a non-linear function of workload size.

\noindent \textit{Energy Predictive Models for Accelerators}. Hong et al. \cite{HongKim2010} present an energy model for an Nvidia GPU based on PMC-based power prediction approach similar to \cite{IcsiMartonosi2003}. Nagasaka et al. \cite{Nagasaka2010} propose PMC-based statistical power consumption modelling technique for GPUs that run CUDA applications. Song et al. \cite{Song2013} present power and energy prediction models based on machine learning algorithms such as backpropagation in artificial neural networks (ANNs). Shao et al. \cite{Shao2013} develop an instruction-level energy consumption model for a Xeon Phi processor.

\textit{Critiques of PMC-based Modelling}. Although PMC based energy predictive software models have become popular in the scientific community, there are several research works which highlight the poor prediction accuracy and limitations of these models. McCullough et al. \cite{McCullough2011} present a study on accuracy of predictive power models for new multicore architectures and show that PMC models based on linear regression gives prediction errors as high as 150\%. O'Brien et al. \cite{Ken2017} survey predictive power and energy models focusing on the highly heterogeneous and hierarchical node architecture in modern HPC computing platforms. They also present an experimental study with linear PMC based energy models where they give an average prediction error equal to 60\%. Economou et al. \cite{Economou2006} highlight the fundamental limitation of PMC-based models, which is the restricted access to read PMCs (generally four at a single run of an application). Shahid et al. \cite{Arsalan2017} propose a selection criterion called the \emph{additivity} for choosing a subset of PMCs to improve the aacuracy of linear energy predictive models. They show that many PMCs in modern multicore CPU platforms fail the \emph{additivity test} and hence are not reliable parameters.

\subsection{Notable Works Involving Performance and Energy}

\cite{Fard2012, Beloglazov2012, Kessaci2013} propose methods for multi-objective optimization involving performance and energy as objectives. Fard et al. \cite{Fard2012} consider four objectives, which are execution time, economic cost, energy, and reliability. Beloglazov et al. \cite{Beloglazov2012} consider twin objectives of energy efficiency and Quality of Service (QoS) for provisioning data center resources. Kessaci et al. \cite{Kessaci2013} present a multi-objective genetic algorithm that minimizes the energy consumption, CO2 emissions and maximizes the generated profit of a cloud computing infrastructure.

\cite{Choi2014, Demmel2013, Drozdowski2014, marszalkowski2016time} are analytical studies of bi-objective optimization for performance and energy. Choi et al. \cite{Choi2014} extend the energy roofline model by adding an extra parameter, power cap, to their execution time model. Drozdowski et al. \cite{Drozdowski2014} use iso-energy map, which are points of equal energy consumption in a multi-dimensional space of system and application parameters, to study performance-energy trade-offs. Marszałkowski et al. \cite{marszalkowski2016time} analyze the impact of memory hierarchies on time-energy trade-off in parallel computations, which are represented as divisible loads. 

The works reviewed in this section do not consider workload distribution as a decision variable.

\section{Formulation of Heterogeneous Dynamic Energy-Performance Optimization Problem (HEPOPT)} \label{sec:formulations}

Consider a workload size $n$ executed using $p$ heterogeneous processors, whose execution time and dynamic energy functions are represented by $T=\{t_0(x),...,t_{p-1}(x)\}$ and $E=\{e_0(x),...,e_{p-1}(x)\}$ where $e_i(x)$ ($t_i(x)$), $i \in \{0, 1, \cdots, p-1\}$, is a discrete dynamic energy (execution time) function with maximum cardinality $m$ for processor $P_i$. The function $e_i(x)$ represents the amount of dynamic energy consumed by $P_i$ to execute the problem size $x$, and $t_i(x)$ is the execution time of the problem size on this processor. Without loss of generality, we assume $x \in \{1, 2, \cdots, m\}$. 

The bi-objective optimization problem to find a workload distribution optimizing execution time and dynamic energy consumption of computations during the parallel execution of workload $n$ using the $p$ processors is formulated as follows:

\begin{equation} \label{eq:bi-obj-de-p}
	\begin{split}
		&HEPOPT(n, p, m, T, E): \min_{X} \quad \{\max_{i=0}^{p-1}~t_i(x_i), \sum_{i=0}^{p-1} e_i(x_i)\}\\
		& \quad \text{Subject to:} \sum_{i=0}^{p-1} x_i = n, 0 \leq x_i \leq m, i \in [0,p-1] \\
		& \quad \text{where} \quad p, n, m \in \mathbb Z_{> 0}, x_i \in \mathbb Z_{\ge 0}, t_i(x), e_i(x) \in \mathbb R_{\ge 0}
	\end{split}
\end{equation}

For each given workload distribution $X = \{x_0,\cdots,x_{p-1}\}$, \emph{HEPOPT} calculates the parallel execution time, which is the time taken by the longest running processor to execute its workload, and the total dynamic energy consumption, which is equal to the summation of dynamic energies consumed by the $p$ processors.

\emph{HEPOPT} returns a set of Pareto-optimal solutions which determine the workload distributions. One or more processors in an optimal solution can be allocated a workload of size zero.

\section{\emph{HEPOPTA}: Algorithm finding Globally Pareto-optimal solutions for execution time and dynamic energy} \label{sec:methodology}

This section illustrates our proposed algorithm, \emph{HEPOPTA} (\textbf{H}eterogeneous \textbf{E}nergy-\textbf{P}erformance \textbf{OPT}imization \textbf{A}lgorithm), solving the problem \emph{HEPOPT}. 

\begin{figure}[!t]
	\centering
	\includegraphics[width=2.5in]{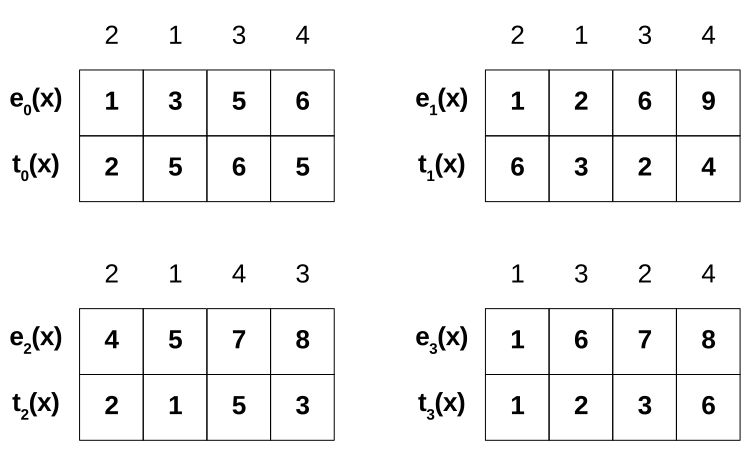}
	\caption{Sample dynamic energy and times functions sorted in non-decreasing order of consumed dynamic energies.}
	\label{fig:sample_functions}
\end{figure}

We describe the algorithm using a simple example. Suppose there are four heterogeneous processors ($p=4$) executing a given workload size $n=4$. The input to \emph{HEPOPTA} are four discrete dynamic energy functions, $E=\{e_0(x),\cdots,e_3(x)\}$, as well as four discrete time functions, $T=\{t_0(x),\cdots,t_3(x)\}$, shown in Figure \ref{fig:sample_functions}. The functions are sorted by dynamic energy in non-decreasing order. They are samples representative of execution time and dynamic energy profiles of real-life data-parallel applications.

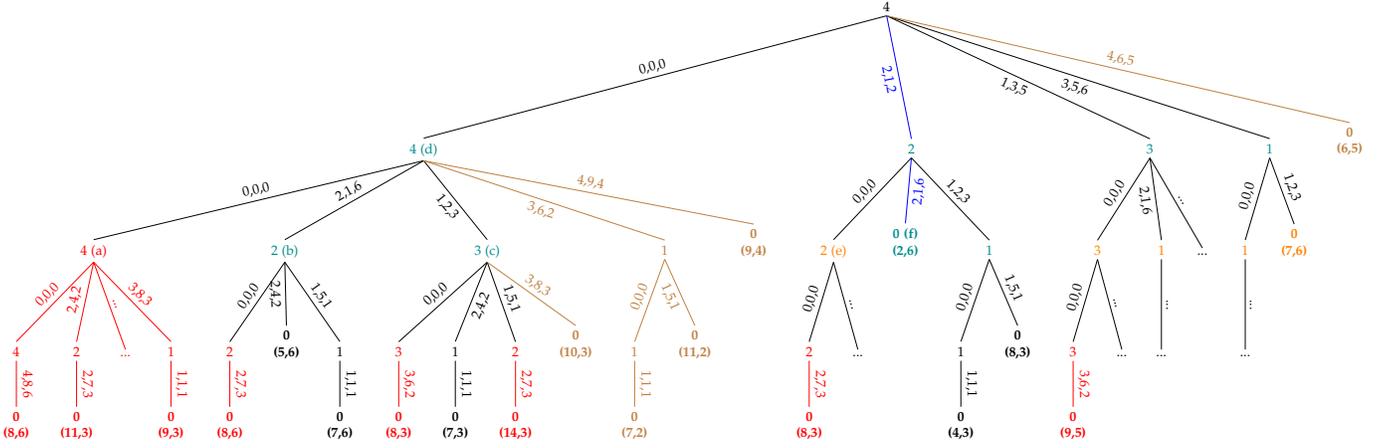
\begin{figure*}[!t]
	\centering
	\begin{tikzpicture} [scale=0.54,sloped]
	\tikzset{level 1/.style={level distance=3.5cm,sibling distance=0.5cm}}
	\tikzset{level 2/.style={level distance=2.5cm,sibling distance=0.5cm}}
	\tikzset{level 3/.style={level distance=2.5cm,sibling distance=0.45cm}}
	\tikzset{level 4/.style={level distance=2cm,sibling distance=0.1cm}}
		\Tree	[.4 \edge node[anchor=south] {0,0,0};
					[.\node[darkcyan] {4 (d)}; \edge node[anchor=south] {0,0,0};
						[.\node[red] {4 (a)}; \edge[red] node[anchor=south] {0,0,0};
							[.\node[red] {4}; \edge[red] node[anchor=south] {4,8,6};
								\node[red] [align=center]{\textbf{0}\\\textbf{(8,6)}};
							]
							\edge[red] node[anchor=south] {2,4,2};
							[.\node[red] {2}; \edge[red] node[anchor=south] {2,7,3};
								\node[red] [align=center]{\textbf{0}\\\textbf{(11,3)}};
							]
							\edge[red] node[anchor=south] {...};
							[ 
								.\node [red]{...};
							]
							\edge[red] node[anchor=south] {3,8,3};
							[.\node[red] {1}; \edge[red] node[anchor=south] {1,1,1};
								\node[red] [align=center]{\textbf{0}\\\textbf{(9,3)}};
							]
						]
						\edge node[anchor=south] {2,1,6};
						[.\node[darkcyan] {2 (b)};	\edge node[anchor=south] {0,0,0};
							[.\node[red] {2}; \edge[red] node[anchor=south] {2,7,3};
								\node [align=center,red]{\textbf{0}\\\textbf{(8,6)}};
							]
							\edge node[anchor=north] {2,4,2};
							[
								.\node [align=center]{\textbf{0}\\\textbf{(5,6)}};
							]
							\edge node[anchor=south] {1,5,1};
							[.1 \edge node[anchor=south] {1,1,1};
								\node [align=center]{\textbf{0}\\\textbf{(7,6)}};
							]
						]
						\edge node[anchor=north] {1,2,3};
						[.\node[darkcyan] {3 (c)}; \edge node[anchor=south] {0,0,0};
							[.\node[red] {3}; \edge[red] node[anchor=south] {3,6,2};
								\node [align=center,red]{\textbf{0}\\\textbf{(8,3)}};						
							]
							\edge node[anchor=north] {2,4,2};
							[.1 \edge node[anchor=south] {1,1,1};
								\node [align=center]{\textbf{0}\\\textbf{(7,3)}};						
							]
							\edge node[anchor=south] {1,5,1};
							[.\node[red] {2}; \edge[red] node[anchor=south] {2,7,3};
								\node [align=center,red]{\textbf{0}\\\textbf{(14,3)}};						
							]
							\edge[brown] node[anchor=south,brown] {3,8,3};
							[
								.\node[align=center,brown]{\textbf{0}\\\textbf{(10,3)}};							
							]					
						]
						\edge[brown] node[anchor=north] {3,6,2};
						[.\node[brown] {1}; \edge[brown] node[anchor=south,brown] {0,0,0};
							[.\node[brown] {1}; \edge[brown] node[anchor=south,brown] {1,1,1};
								\node[brown] [align=center,brown]{\textbf{0}\\\textbf{(7,2)}};						
							]
							\edge[brown] node[anchor=north,brown] {1,5,1};
							[
								.\node [align=center,brown]{\textbf{0}\\\textbf{(11,2)}};						
							]							
						]	
						\edge[brown] node[anchor=south,brown] {4,9,4};
						[
							.\node [align=center,brown]{\textbf{0}\\\textbf{(9,4)}};
						]
					]
					\edge[blue] node[anchor=north] {2,1,2};
					[.\node[darkcyan] {2};  \edge node[anchor=south] {0,0,0};
						[.\node[orange] {2 (e)}; \edge node[anchor=south] {0,0,0};
							[.\node[red]{2}; \edge[red] node[anchor=south] {2,7,3};
								\node [align=center,red]{\textbf{0}\\\textbf{(8,3)}};
							]
							\edge node[anchor=south] {...};
							[
								.\node {...};
							]
						]
						\edge[blue] node[anchor=north] {2,1,6};
						[
							.\node [align=center,color=darkcyan]{\textbf{0 (f)}\\\textbf{(2,6)}};
						]						
						\edge node[anchor=south] {1,2,3};
						[.\node[darkcyan] {1}; \edge node[anchor=south] {0,0,0};
							[.1 \edge node[anchor=south] {1,1,1};
								\node [align=center]{\textbf{0}\\\textbf{(4,3)}};
							]
							\edge node[anchor=south] {1,5,1};
							[
								.\node [align=center]{\textbf{0}\\\textbf{(8,3)}};
							]
						]
					]
					\edge node[anchor=north] {1,3,5};
					[.\node[darkcyan] {3}; \edge node[anchor=south] {0,0,0};
						[.\node[orange] {3}; \edge node[anchor=south] {0,0,0};
							[.\node[red] {3}; \edge[red] node[anchor=south] {3,6,2};
								\node [align=center,red]{\textbf{0}\\\textbf{(9,5)}};
							]
							\edge node[anchor=south] {...};
							[
								.\node {...};
							]
						]
						\edge node[anchor=north] {2,1,6};
						[.\node[orange] {1}; \edge node[anchor=south] {...};
							[
								.\node {...};
							]
						]
						\edge node[anchor=south] {...};
						[
							.\node {...};
						]
					]
					\edge node[anchor=north] {3,5,6};
					[.\node[darkcyan] {1}; \edge node[anchor=south] {0,0,0};
						[.\node[orange] {1}; \edge node[anchor=south] {...};
							[
								.\node {...};
							]				
						]
						\edge node[anchor=south] {1,2,3};
						[
							.\node [align=center,orange]{\textbf{0}\\\textbf{(7,6)}};				
						]
					]			
					\edge[brown] node[anchor=south,brown] {4,6,5};
					[
						.\node [align=center,brown]{\textbf{0}\\\textbf{(6,5)}};
					]
				]		
	\end{tikzpicture}
	\caption{The solution tree explored by the naive algorithm to find all distributions and its Pareto-optimal set for a workload $n = 4$ on four processors.}
	\label{fig:search_tree}
\end{figure*}

To find the Pareto-optimal solutions for execution time and dynamic energy, a straightforward approach is to explore the full solution tree and find all possible workload distributions. Figure \ref{fig:search_tree} shows the tree, which is constructed by such a naive algorithm. Due to the lack of space, we only show the tree partially. 

The tree consist of $4$ levels $\{L_0,L_1,L_2,L_3\}$ where all problem sizes given to processor $P_i$ are examined in level $L_i$. Each node in $L_i$, $i \in \{0,1,2,3\}$, is labelled by a positive value representing the workload size that is distributed between processors $\{P_i,\cdots,P_3\}$. Each edge connecting a node at level $L_i$ to its ancestor is labelled by a triple $(w,e,t)$ where $w$ is the problem size assigned to $P_i$, along with its consumed dynamic energy ($e_i(w)$) and its execution time ($t_i(w)$).

The exploration process begins from the root to find all distributions for the workload size four between four processors $\{P_0, P_1, P_2, P_3\}$. Five problem sizes, including all data points in the function $e_0(x)$ and a zero problem size, are assigned to the processor $P_0$ one after another. Although there is no ordering assumption, we examine the problem sizes in this example in non-decreasing order of their dynamic energy consumption. Assigning the problem sizes $\{0,2,1,3,4\}$ to $P_0$ expands the root into 5 children at $L_1$ representing the remaining workload to be distributed between processors $\{P_1, P_2, P_3\}$. For instance, the edge $(2,1,2)$, highlighted in blue in Figure \ref{fig:search_tree}, indicates that a problem size $2$ with a dynamic energy consumption of $1$ and an execution time of $2$ is given to $P_0$, and its child is labelled by $2$ which equals the remaining size distributed at the level $L_1$. 

In the same manner, each node in levels $\{L_1,L_2,L_3\}$ are expanded towards the leaves. Any leaf node, labelled by $0$, illustrates a solution that its dynamic energy consumption is the summation of dynamic energy consumptions and its execution time is the maximum execution times labelling the edges in the path from the root to the leaf. For example, the blue path $\{(2,1,2),(2,1,6)\}$ in the tree highlights a solution distributing the workload $4$ on two processors $P_0$ and $P_1$ where its dynamic energy consumption is $2~(= 1 + 1)$, and its execution time equals $6 =~(\max \{2,6\})$. It is obvious that the other two processors $\{P_2,P_3\}$ are assigned a zero problem size.

Due to lack of space, we have not shown the branches that do not provide any solution. In a non-solution branch, the summation of problem sizes labelling the edges from the root to its leaf is greater than $4$.

In this example, each internal node in the solution tree has either $5$ children (or $m+1$ in general case) or just one child in which case the child is always a leaf. There are two types of leaves: \emph{solution} leaves, labelled by $0$ along with its dynamic energy consumption and execution time beneath it, and \emph{no-solution} leaves, eliminated from, and therefore, not shown in the tree. Each internal node at level $L_i$, labelled by a positive number $w$, becomes a root of a solution tree for distribution of the workload $w$ between processors $\{P_i, \cdots, P_3\}$ and is therefore constructed recursively.  

Once a solution is found, the algorithm updates the Pareto-optimal set. In the end, the globally Pareto-optimal set includes three members, $\{(\langle2,6\rangle,\{2,2,0,0\}), (\langle4,3\rangle,\{2,1,0,1\}), (\langle5,2\rangle,\{2,0,2,0\})\}$, where each element, like $(\langle eng, eTime \rangle,\{x_0,\cdots,x_3\})$, in the set determines the dynamic energy consumption ($eng$) and the execution time ($eTime$) of the workload distribution $\{x_0,\cdots,x_3\}$.

The naive algorithm has exponential complexity. We propose \emph{HEPOPTA} which is an efficient recursive algorithm to determine the globally Pareto-optimal set of solutions for data-parallel applications executing on heterogeneous processors. It has polynomial computational complexity. The algorithm shrinks the search space by utilizing three optimizations to avoid exploring whole subtrees in the solution tree.

\begin{figure}[!t]
	\centering
	\includegraphics[width=2.5in]{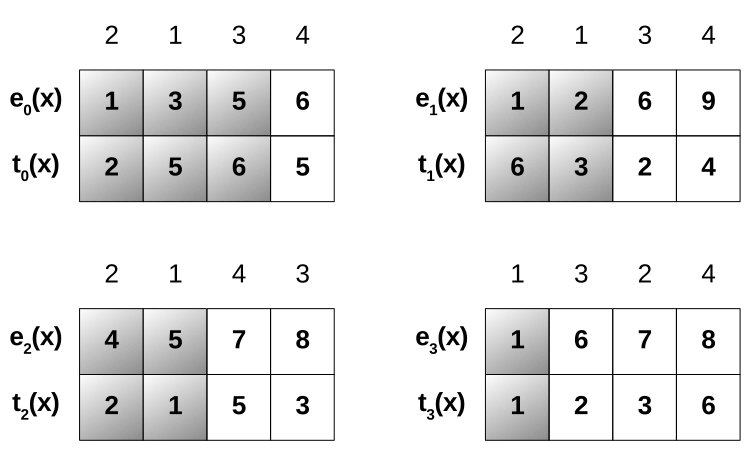}
	\caption{Removing some data points from the profiles by applying the energy threshold $\varepsilon$.}
	\label{fig:sample_functions_highlited}
\end{figure}

We will now explain how \emph{HEPOPTA} efficiently solves the aforementioned example. It scans dynamic energy functions, starting with $e_0(x)$, from left to right in non-decreasing order of dynamic energy consumption. The first optimization concerns the upper bound for dynamic energy consumption, which we call it \emph{energy threshold} represented by $\varepsilon$. It is the dynamic energy consumption of the workload distribution which optimizes the execution time of the workload $4$ on the processors. We determine this optimal distribution by using the algorithm \emph{HPOPTA} \cite{khaleghzadeh2018novel}, which finds optimal workload distribution minimizing the execution time. We then initialize the variable $\varepsilon$ to the dynamic energy consumption of this distribution. Applying energy threshold enables \emph{HEPOPTA} to shrink search space by ignoring all data points with consumed dynamic energies greater than $\varepsilon$. In the example, the optimal workload distribution, returned by \emph{HPOPTA}, is $X_{t_{opt}} = \{2,0,2,0\}$ with an execution time ($t_{opt}$) of $2$. Therefore, $\varepsilon$ in this example is set to $5$, which is the dynamic energy consumption for this distribution ($\sum_{i=0}^{p-1}e_i(x_{t_{opt}}[i])=5$). \emph{HEPOPTA}, as shown in Figure \ref{fig:sample_functions_highlited}, ignores all data points whose dynamic energy consumptions are greater than $5$. We highlight in brown all nodes and branches eliminated from the solution tree by deploying energy threshold in Figure \ref{fig:search_tree}. There may exist more than one workload distribution minimizing the execution time but with different dynamic energy consumptions. It is obvious that the best solution is the distribution which minimizes $\varepsilon$. Nevertheless, using a non-optimal $\varepsilon$ does not restrain \emph{HEPOPTA} from obtaining the globally Pareto-optimal set.

To shrink the search space further, \emph{HEPOPTA} assigns each level of the tree a size threshold $\sigma_i, i \in \{0,\dots,p-1\}$. It represents the maximum workload which can be executed in parallel on processors $\{P_i,\cdots,P_{p-1}\}$ so that the dynamic energy consumption of each processor in $\{P_i,\cdots,P_{p-1}\}$ is not greater than $\varepsilon$. In this example, the size threshold vector $\sigma$ contains four elements, $\sigma = \{\sigma_0,\sigma_1,\sigma_2,\sigma_3\} = \{8,5,3,1\}$. Before expanding each node, \emph{HEPOPTA} compares its workload with its corresponding size threshold. If the workload exceeds the size threshold, the node is not expanded since it results in a solution with a dynamic energy consumption greater than $\varepsilon$. 

After calculating the energy threshold $\varepsilon$ and the size threshold vector $\sigma$, \emph{HEPOPTA} explores the solution tree from its root in the left-to-right and depth-first order. It, first, allocates zero problem sizes to $P_0$ and $P_1$ (Figure \ref{fig:search_tree}). The remaining workload at the level $L_2$ is $4$ which is labelled by $4 (a)$ in the tree. Since the workload $4$ is greater than the corresponding size threshold $\sigma_2$, the node is not expanded further and is cut. This optimization is called operation \emph{Cut}. We highlight in red all sub-trees eliminated from the search space using the operation \emph{Cut}.

Returning to the tree exploration, \emph{HEPOPTA} examines the next node $2 (b)$ at the level $L_2$. Expansion of this node results in two solutions partitioning workload $2$ on processors $P_2$ and $P_3$. \emph{HEPOPTA} updates the Pareto-optimal set for this node and saves the solution in memory called $PMem$. 

\emph{HEPOPTA} memorizes solutions for each node in levels $\{L_1,\cdots,L_{p-2}\}$. The information stored for a node with a workload of $w$ at a given level $L_i$, $i \in \{1,\cdots,p-2\}$, is a quintuple $<eng,time, part, P\#, key>$ where $eng$ is the dynamic energy consumption of the solution, $time$ is its parallel execution time on processors $\{P_i,\cdots,P_{p-1}\}$, $part$ is the problem size given to $P_i$, $P\#$ is the number of active processors in the solution and finally, $key$, is set to the dynamic energy consumption of a saved Pareto-optimal solution for workload $w-c$ at level $L_{i+1}$. We call this Pareto-optimal solution at level $L_{i+1}$ a \textit{partial solution} for the workload $w$. This partial solution may not exist for some nodes, where in this case we represent it by $\emptyset$. Since dynamic energies are unique in a Pareto-optimal set, we use $key$ as a pointer to partial solutions. For each solution leaf in levels $\{L_1,\cdots,L_{p-2}\}$, like $0 (f)$ in Figure \ref{fig:search_tree}, \emph{HEPOPTA} memorizes a solution $\{<0,0,0,0,\emptyset>\}$.

Thus, the information saved for the node $2 (b)$ is a Pareto-optimal set including two members, $\{<4,2,2,1,\emptyset>,<6,1,1,2,\emptyset>\}$. We call this key operation, \emph{SavePareto}. Green nodes in the solution tree highlight ones whose Pareto-optimal sets are saved. After $2 (b)$, the node $3 (c)$ is examined. The solution saved for this node is $\{<5,2,2,2,\emptyset>\}$. 

\emph{HEPOPTA} then backtracks to the node $4 (d)$ on $L_1$ and builds its Pareto-optimal set by merging Pareto-optimal sets saved for its children, $2 (b)$ and  $3 (c)$. Consider the edge $(2,1,6)$ connecting the node $4 (d)$ to $2 (b)$. Merging this edge with the Pareto-optimal set which has been already saved for $2 (b)$, $\{<4,2,2,1,\emptyset>,<6,1,1,2,\emptyset>\}$, results in one Pareto-optimal solution for the node $4 (d)$ which is saved as the quintuple $<5,6,2,2,\textbf{4}>$. In this solution, the last element , $4$, which is highlighted in bold, points to its partial solution in the node $2 (b)$ at $L_2$, which is $\{<4,2,2,1,\emptyset>\}$. Merging the edge $(1,2,3)$ with the Pareto-optimal set for $3 (c)$, $\{<5,2,2,2,\emptyset>\}$, results in a new solution $\{<7,3,1,3,5>\}$. Therefore, the Pareto-optimal set for the node $4 (d)$ is $\{<7,3,1,3,5>, <5,6,2,2,4>\}$, which is saved in the memory. 

After building and saving the Pareto-optimal set of the node $4 (d)$, \emph{HEPOPTA} visits the node $2 (e)$ at the level $L_2$. This node has already been explored, and therefore, its Pareto-optimal set is retrieved from $PMem$. We call this key operation, \emph{ReadParetoMem}. The nodes whose solutions are retrieved from the memory are highlighted in orange.

After visiting the other remaining nodes, \emph{HEPOPTA} backtracks to the root and builds the globally Pareto-optimal solutions for the workload $4$ executing on processors $\{P_0,\cdots,P_3\}$ using the Pareto-optimal sets saved for its children. Then it terminates.

\emph{HEPOPTA} thus deploys three key operations, which are a). \emph{Cut}, b). \emph{SavePareto}, and c). \emph{ReadParetoMem}, to efficiently explore solution trees and build globally Pareto-optimal solutions optimizing for execution time and dynamic energy.

\section{Formal Description of \textit{HEPOPTA}}	\label{sec:HEPOPTAformal}

We present the pseudocode of \emph{HEPOPTA} in Algorithm \ref{alg_code_HEPOPTA}. The Inputs of the algorithm are: the problem size, $n$, the number of heterogeneous processors, $p$, an array of $p$ dynamic energy profiles, $E=\{E_0,E_1,\cdots,E_{p-1}\}$ and $p$ time functions $T=\{T_0,T_1,\cdots,T_{p-1}\}$ where $E_i$ is the dynamic energy function, and $T_i$ represents the execution time of processor $P_i$, $i \in \{0,\cdots,p-1\}$. Each energy function comprises $m$ pairs $(x_{ij},e_{ij})$, $j \in \{0,1,\cdots,m-1\}$, so that $x_{ij}$ is the j-th problem size in the function and $e_{ij}$ represents the amount of dynamic energy consumed by running it on $P_i$. Each time function includes $m$ pairs $(x_{ij},t_{ij})$, $j \in \{0,1,\cdots,m-1\}$, so that $x_{ij}$ is the j-th problem size in the function and $t_{ij}$ represents its execution time on $P_i$. \emph{HEPOPTA} returns $\Psi_{EP}$, the globally Pareto-optimal solutions. It consists of a set where each element of it is a triple like $(eng, time, X)$. The first field $eng$ is the dynamic energy consumption of a Pareto-optimal solution, $time$ represents its execution time, and $X=\{x_0, x_1, \cdots, x_{p-1}\}$ determines the workload distribution of the solution. The solutions are sorted in increasing order of dynamic energy.

\emph{HEPOPTA} starts by sorting energy and time functions in non-decreasing order of dynamic energy consumption and execution time (Line \ref{hepopta_sorting1}). Both original and sorted functions are kept. Original functions are assumed to be sorted by problem size. Then, $\Call{HPOPTA}{}$ \cite{khaleghzadeh2018novel} is invoked to find the optimal distribution minimizing the execution time of the workload $n$ on $p$ processors (Line \ref{hepopta_optE_time}). This function returns the optimal execution time, $t_{opt}$, along with its distribution, $X_{t_{opt}}$. The energy threshold $\varepsilon$ is initialised to the dynamic energy consumption of the distribution $X_{t_{opt}}$ (Line \ref{hepopta_e-threshold}). The function \Call{ReadFunc}{$E_i,x$} returns the dynamic energy consumption of the problem size $x$ executing on the processor $P_i$. It returns $0$ when $x$ is equal to $0$. 

The size threshold array $\sigma$ is initialised by using the function \Call{SizeThresholdCalc}{} (Line \ref{hepopta_s_th}). A 2D array $PMem$, with dimensions of $(p-2) \times (n+1)$, is defined to save Pareto-optimal solutions for processors $\{P_1,\cdots,P_{p-2}\}$, which are found during the tree exploration (Line \ref{hepopta_mem}). Then, \Call{HEPOPTA\_Kernel}{} is invoked to explore the solution tree and determines the globally Pareto-optimal set of solutions for execution time and dynamic energy, returned in $\Psi_{EP}$.

The pseudocodes of all functions, including \Call {HEPOPTA\_Kernel}{}, its correctness and complexity proofs, and the structure of $PMem$ are described in Section \ref{sec:append}. 

\begin{algorithm}
	\scriptsize
	\caption{Algorithm finding globally Pareto-optimal solutions for execution time and dynamic energy of a workload $n$ on $p$ heterogeneous processors.} \label{alg_code_HEPOPTA}
	\begin{algorithmic}[1]	
		\Function{HEPOPTA}{$n, p, E, T, \Psi_{EP}$}
		\Statex \textbf{INPUT:}
		\Statex Problem size, $n \in \mathbb Z_{> 0}$
		\Statex Number of processors, $p \in \mathbb Z_{> 0}$
		\Statex Dynamic energy profiles, $E = \{E_0,...,E_{p - 1}\}$,
		\Statex $E_i = \{(x_{ij},e_{ij})~|~i \in [0,p), j \in [0,m), x_{ij} \in \mathbb Z_{> 0}, e_{ij} \in \mathbb R_{> 0}\}$.
		\Statex Time functions, $T = \{T_0,...,T_{p - 1}\}$,
		\Statex $T_i = \{(x_{ij},t_{ij})~|~i \in [0,p), j \in [0,m), x_{ij} \in \mathbb Z_{> 0}, t_{ij} \in \mathbb R_{> 0}\}$.
		\Statex \textbf{OUTPUT:}
		\Statex Pareto-optimal solutions for execution time and dynamic energy, $\Psi_{EP}$,
		\Statex $\Psi_{EP} = \{(eng_k,time_k,X_k)~|~k \in [0,|\Psi_{EP}|)\}$,
		\Statex $X_k = \{x_k[0],x_k[1],\cdots,x_k[p-1]\}$,
		\Statex $x_k[i] \in \{\bigcup_{j = 0}^{m-1} x_{ij} \cup \{0\}\},~i \in [0,p)$.
		\Statex
		\State $E$ $\gets$ $E \cup Sort_\uparrow(E)$ $, $ $T$ $\gets$ $T \cup Sort_\uparrow(T)$	\label{hepopta_sorting1}
		
		\State $(X_{t_{opt}}, t_{opt})$ $\gets$ \Call {HPOPTA}{$n, p, T$}		\label{hepopta_optE_time}
		\State $\varepsilon$ $\gets$ $\sum_{i=0}^{p-1} \Call {ReadFunc} {E_i, x_{t_{opt}}[i]}$	\label{hepopta_e-threshold}
		\State $\sigma$ $\gets$ \Call {SizeThresholdCalc}{$p,E,\varepsilon$}		\label{hepopta_s_th}
		\State $PMem[i][j]$ $\gets$ $\emptyset$, $\forall i \in \{1,\cdots,p-2\},$ $j \in \{0,\cdots,n\}$ \label{hepopta_mem}
		\State \Call {HEPOPTA\_Kernel}{$n, p, 0, E, T, \varepsilon, \sigma, X_{cur}, PMem, \Psi_{EP}$} \label{hepopta_kernelCall}
		\State \Return $\Psi_{EP}$
		\EndFunction		
	\end{algorithmic}
\end{algorithm}

\subsection{Recursive Algorithm $HEPOPTA\_Kernel$} \label{sec_code_heopta}

Algorithm \ref{hep_kernel_code} shows the pseudocode for \textit{HEPOPTA\_Kernel}. It efficiently explores the solution tree and recursively builds Pareto-optimal solutions from tree leaves to the root. Pareto-optimal solutions for a given node at level $L_i$, $i \in \{0,1,\cdots,p-2\}$, are built by merging all solutions stored for its children, placed at level $L_{i + 1}$. \textit{HEPOPTA\_Kernel} uses three operations \emph{Cut}, \emph{SavePareto} and \emph{ReadParetoMem}, illustrated in the main manuscript, to reduce the search space and achieve a polynomial computational complexity.

The variable $c \in \{0,\cdots,p-1\}$ indicates the tree level that is processing in the current recursion of \textit{HEPOPTA\_Kernel}. Prior to expanding a node at the level $L_c$, \textit{HEPOPTA\_Kernel} determines whether its workload exceeds $\sigma_{c}$. If it is the case then the node is not explored (Lines \ref{hep_kernel_size_thr1}-\ref{hep_kernel_size_thr2}). Lines \ref{hep_kernel_leaf1}-\ref{hep_kernel_leaf2} process solutions found at the last level $L_{p-1}$. If there exists a solution, the function returns $TRUE$, otherwise $FALSE$. 

Before exploring a node at a given level $c$, $c \in \{L_1,\cdots,L_{p-2}\}$, the function \Call{ReadParetoMem}{} is called to retrieve from $PMem$ the solution set saved for the current workload $n$ on processors $\{P_c,\cdots,P_{p-1}\}$ (Lines \ref{hep_kernel_retrieveMem1}-\ref{hep_kernel_retrieveMem2}). The variable \emph{status} determines the type of retrieved solutions. If no solution is already stored for the node or the total dynamic energy consumption of all the retrieved solutions is greater than or equal to $\varepsilon$ (given by the status, \emph{NOT\_SOLUTION}), \emph{HEPOPTA\_Kernel} returns $FALSE$ and backtracks. If at least one of the solutions, in the retrieved set, has a total dynamic energy consumption less than $\varepsilon$ (given by the status, \emph{SOLUTION}), the function returns $TRUE$. If none of the above cases happen, the routine starts expanding the node by initializing pointer $idx$ to $-1$ and $x_{c~idx}$ to $0$ (Lines \ref{hep_kernel_setzero}-\ref{hep_kernel_mainLoop2}). The variable $idx$, ranging from $-1$ to $m-1$, determines indexes of data points in the functions, and $x_{c~idx}$ represents the problem size of $idx$-th data point in the functions.

The $while$ loop (Lines \ref{hep_kernel_mainLoop1}-\ref{hep_kernel_mainLoop2}) examines all data points with dynamic energy consumption less than or equal to $\varepsilon$ in the function $E_c$, sorted in non-decreasing order of energy consumption. The array $X_{cur}=\{x_{cur}[0],\cdots,x_{cur}[p-1]\}$, where $x_{cur}[i] \in \{\bigcup_{j = 0}^{m-1} x_{ij} \cup \{0\}\} $, stores problem sizes currently assigned to processors $P_i (i \in \{0,1,\cdots,p-1\}$). In each iteration, the data point $idx$ is extracted from $E_c$, and its problem size ($x_{c~idx}$) is stored in array $x_{cur}[c]$ (Line \ref{hep_kernel_store_dc}). \emph{HEPOPTA\_Kernel} is recursively invoked to find solutions for the remaining workload $n-x_{c~idx}$ at the next level $L_{c+1}$ (Line \ref{hep_kernel_recall}). If there exists any solution for the workload, $x_{c~idx}$ is added to $partsVec$, a list holding all problem sizes, given to $P_c$, which result in Pareto-optimal solutions (Lines \ref{hep_kernel_onreturn1}-\ref{hep_kernel_onreturn2}). 

If $idx$ reaches the end of the energy profile $E_c$, the \emph{while} loop terminates (Lines \ref{hep_kernel_break1}-\ref{hep_kernel_break2}), otherwise, $idx$ is incremented to examine the next data point in the energy profile $E_c$.

After exploring all children of the current node, the function \Call{MergePartialParetoes}{} is invoked to merge and store the Pareto-optimal solutions of its children into a single Pareto-optimal set of solutions. 

In the end, the corresponding memory cell storing the Pareto-optimal solution for a node with a workload $n$ at $L_c$ ($PMem[c][n]$) is labelled \emph{Finalized} (Line \ref{hep_kernel_fin}). Finalizing a memory cell implies that this cell contains the final solutions. \textit{HEPOPTA\_Kernel} returns $TRUE$ provided that exploring the node, processed in the current recursion, has led to a solution (Line \ref{hep_kernel_return}).

\begin{algorithm}
	\scriptsize
	\caption{Recursive Kernel Invoked by Function $HEPOPTA$} \label{hep_kernel_code}
	\begin{algorithmic}[1]	
		\Function{HEPOPTA\_Kernel}{$n, p, c, E, T, \varepsilon, \sigma, X_{cur}, PMem, \Psi_{EP}$}
		\Statex
			\If{\Call{Cut}{$n, \sigma_{c}$}}					\label{hep_kernel_size_thr1}
				\State \Return $FALSE$
			\EndIf												\label{hep_kernel_size_thr2}
			\If{$c = p - 1$ $\wedge$ \Call{ReadFunc}{$E_c,n$} $ \le \varepsilon$}		\label{hep_kernel_leaf1}
				\State $x_{cur}[c]$ $\gets$ $n$
				\State \Return $TRUE$
			\Else
				\State \Return $FALSE$
			\EndIf									\label{hep_kernel_leaf2}
			\If{$n \neq 0$ $\wedge$ $c \ge 1$ $\wedge$ $c \leq p - 2$}	\label{hep_kernel_retrieveMem1}
				\State $status$ $\gets$ \Call{ReadParetoMem}{$n, c, \varepsilon, PMem$} \label{hep_kernel_rSolMem}
				\If{$status=NOT\_SOLUTION$}
					\State \Return $FALSE$
				\ElsIf{$status=SOLUTION$}		
					\State \Return $TRUE$
				\EndIf
			\EndIf									\label{hep_kernel_retrieveMem2}
			\State $idx$ $\gets$ $-1$ $; $ $x_{c~idx}$ $\gets$ $0$	\label{hep_kernel_setzero}
			\State $isSol$ $\gets$ $FALSE$
			\State $partsVec \gets \varnothing$
			\While{\Call{ReadFunc}{$E_c,x_{c~idx}$} $ \le \varepsilon$} 			\label{hep_kernel_mainLoop1}
				\If{$x_{c~idx} \le n$}			
					\State $x_{cur}[c]$ $\gets$ $x_{c~idx}$	\label{hep_kernel_store_dc}									
					\State $outRes$ $\gets$ \Call{HEPOPTA\_Kernel}{$n-x_{c~idx},p,c+1,E, T, \varepsilon,\sigma,X_{cur}, PMem, \Psi_{EP}$}	\label{hep_kernel_recall}
					\If{$outRes = TRUE$}	\label{hep_kernel_onreturn1}
						\State $isSol$ $\gets$ $TRUE$
						\State $partsVec$ $\gets$ $partsVec$ $\cup$ $x_{c~idx}$
					\EndIf					\label{hep_kernel_onreturn2}
				\EndIf				
				\If{$n = 0$ $\vee$ $idx+1=m$}				\label{hep_kernel_break1}
					\State \textbf{break}	
				\EndIf						\label{hep_kernel_break2}
				\State $idx \gets idx+1$	\label{hep_kernel_nextpoint}
			\EndWhile								\label{hep_kernel_mainLoop2}
			\If{$c \ge 1$ $\wedge$ $c \leq p - 2$}	\label{hep_kernel_merge1}
				\State \Call {MergePartialParetoes}{$n, p, c, E, T,partsVec,PMem,\Psi_{EP}$}	
			\EndIf									\label{hep_kernel_merge2}
			\State \Call{MakeParetoFinal}{$PMem[c][n]$}	\label{hep_kernel_fin}
			\State \Return $isSol$	\label{hep_kernel_return}			
		\EndFunction		
	\end{algorithmic}
\end{algorithm}

\section{\emph{HTPOPTA}: Algorithm finding Globally Pareto-optimal solutions for performance and total energy} \label{sec:total-eng-optimiztion}

Consider a workload size $n$ executing using $p$ heterogeneous processors, whose execution time and dynamic energy functions are represented by $T=\{t_0(x),...,t_{p-1}(x)\}$ and $E=\{e_0(x),...,e_{p-1}(x)\}$, and $P_S$ is the base power of the platform.

The bi-objective optimization problem for performance and total energy, \emph{HTPOPT}, to obtain workload distributions minimizing execution time and total energy consumption during the parallel execution of the workload $n$ using the $p$ processors can be formulated as follows:

\begin{equation} \label{eq:bi-obj-te-p}
	\begin{split}
		&HTPOPT(n, p, m, T, E, P_S): \\
		& \qquad \min_{X} \quad \{\max_{i=0}^{p-1}~t_i(x_i), P_S \times \max_{i=0}^{p-1}~t_i(x_i) + \sum_{i=0}^{p-1} e_i(x_i)\}\\
		& \quad \text{Subject to:} \sum_{i=0}^{p-1} x_i = n, 0 \leq x_i \leq m, i \in [0,p-1] \\
		& \text{where} \quad p, n, m \in \mathbb Z_{> 0}, x_i \in \mathbb Z_{\ge 0}, t_i(x), e_i(x), P_S \in \mathbb R_{\ge 0}
	\end{split}
\end{equation}

We prove that the solution to the problem \emph{HTPOPT} is a subset of the globally Pareto-optimal set of solutions for execution time and dynamic energy determined by the algorithm \emph{HEPOPTA}. The correctness proof is presented in Section \ref{sec:append}.

We propose an algorithm called \emph{HTPOPTA} (\textbf{H}eterogeneous \textbf{T}otal energy-\textbf{P}erformance \textbf{OPT}imization \textbf{A}lgorithm) solving \emph{HTPOPT}. It takes as inputs, the workload size, $n$; the number of heterogeneous processors, $p$; an array of $p$ dynamic energy profiles, $E=\{E_0,E_1,\cdots,E_{p-1}\}$; an array of $p$ time functions $T=\{T_0,T_1,\cdots,T_{p-1}\}$; and the base power of the platform ($P_{S}$). It returns the globally Pareto-optimal solutions for execution time and total energy. \emph{HTPOPTA} calls \emph{HEPOPTA} to find the solutions.

\subsection{Formal Description of \textit{HTPOPTA}}
	
The function \emph{HTPOPTA} calculates globally Pareto-optimal solutions for total energy and performance using $\Psi_{EP}$. It takes as input the problem size, $n$, the number of heterogeneous processors, $p$, an array of $p$ dynamic energy functions, $E=\{E_0,E_1,\cdots,E_{p-1}\}$, an array of $p$ time functions $T=\{T_0,T_1,\cdots,T_{p-1}\}$ and the base power of the platform, $P_{S}$. \emph{HTPOPTA} returns the globally Pareto-optimal set for execution time and total energy which are stored in $\Psi_{TP}$. It is a set of triples like $(teng, time, X)$ where $teng$ illustrates the total energy consumption of a Pareto-optimal solution, $time$ is its execution time, and $X=\{x_0, x_1, \cdots, x_{p-1}\}$ represents the workload distribution of the solution.

\emph{HTPOPTA}, first, calls \Call{HEPOPTA}{} to find globally Pareto-optimal solutions for dynamic energy and performance, $\Psi_{EP}$ (Line \ref{TPPareto_HEPOPTA}). It then calculates the total energy consumption of every solution in $\Psi_{EP}$ (Line \ref{TPPareto_te}) and enquiries if there exists a solution in $\Psi_{TP}$ where its total energy consumption is equal to that of the new solution but with less execution time or with the same execution times but less active processors. If this is the case, the current solution in $\Psi_{TP}$ is updated by the new one (Lines \ref{TPPareto_update1}-\ref{TPPareto_update2}). Otherwise, the new solution is added into $\Psi_{TP}$ (Line \ref{TPPareto_add}).

After inserting solutions, non-Pareto-optimal solutions are found (Lines \ref{TPPareto_eliminate1}-\ref{TPPareto_eliminate2}) to get eliminated from $\Psi_{TP}$ (Line \ref{TPPareto_return}). Pareto-optimal solutions in $\Psi_{TP}$ are also sorted in the increasing order of total energy consumption and decreasing order of execution time. It should be noted that solutions in $\Psi_{EP}$ and $\Psi_{TP}$ are sorted in increasing order of energy consumption, and consequently in decreasing order of execution time.

\begin{algorithm}
	\scriptsize
	\caption{Algorithm Finding Globally Pareto-optimal Solutions for Total Energy and Performance using \emph{HEPOPTA}} \label{alg_TPparetoSols}
	\begin{algorithmic}[1]	
		\Function{$HTPOPTA$}{$n, p, E, T, P_S, \Psi_{TP}$}
			\Statex \textbf{INPUT:}
			\Statex Problem size, $n \in \mathbb Z_{> 0}$
			\Statex Number of processors, $p \in \mathbb Z_{> 0}$
			\Statex Energy profiles, $E = \{E_0,...,E_{p - 1}\}$,
			\Statex $E_i = \{(x_{ij},e_{ij})~|~i \in [0,p), j \in [0,m), x_{ij} \in \mathbb Z_{> 0}, e_{ij} \in \mathbb R_{> 0}\}$.
			\Statex Time functions, $T = \{T_0,...,T_{p - 1}\}$,
			\Statex $T_i = \{(x_{ij},t_{ij})~|~i \in [0,p), j \in [0,m), x_{ij} \in \mathbb Z_{> 0}, t_{ij} \in \mathbb R_{> 0}\}$.
			\Statex Base power of the heterogeneous platform, $P_S \in \mathbb R_{> 0}$
			\Statex \textbf{OUTPUT:}
			\Statex Pareto-optimal solutions for total energy and performance, $\Psi_{TP}$,			
			\Statex $\Psi_{TP} = \{(teng_k,time_k,X_k)~|~k \in [0,|\Psi_{TP}|)\}$,
			\Statex $X_k = \{x_k[0],x_k[1],\cdots,x_k[p-1]\}$,
			\Statex $x_k[i] \in \{\bigcup_{j = 0}^{m-1} x_{ij} \cup \{0\}\},~i \in [0,p)$.
			\Statex
			\State \Call{HEPOPTA}{$n, p, E, T, \Psi_{EP}$}	\label{TPPareto_HEPOPTA}
			\ForAll{$tup \in \Psi_{EP}$}
				\State $te \gets tup.eng + P_S \times tup.time$	\label{TPPareto_te}
				\State $tup' \gets \{x~|~x \in \Psi_{TP}, x.eng = te\}$		\label{TPPareto_update1}
				\If{$tup' \neq \emptyset$}			
					\If{$tup.time < tup'.time$}
						\State $tup' \gets (te,tup.time,tup.X)$
					\ElsIf{$tup.time = tup'.time$}
						\State $idle_{tup} \gets 0$
						\State $idle_{tup'} \gets 0$
						\ForAll{$x \in tup.X$}
							\If{$x=0$}
								\State $idle_{tup} \gets idle_{tup} + 1$
							\EndIf
						\EndFor
						\ForAll{$x \in tup'.X$}
							\If{$x=0$}
								\State $idle_{tup'} \gets idle_{tup'} + 1$
							\EndIf
						\EndFor
						\If{$idle_{tup} < idle_{tup'}$}
							\State $tup' \gets (te,tup.time,tup.X)$
						\EndIf
					\EndIf											\label{TPPareto_update2}
				\Else
					\State $\Psi_{TP} \gets \Psi_{TP} \cup (te,tup.time,tup.X)$	\label{TPPareto_add}
				\EndIf
			\EndFor
			\State $minTime \gets \infty$
			\State $nPList \gets \emptyset$
			\ForAll{$tup \in \Psi_{TP}$}			\label{TPPareto_eliminate1}
				\If{$tup.time \ge minTime$}
					\State $nPList \gets nPList \cup tup$
				\Else
					\State $minTime \gets tup.time$
				\EndIf
			\EndFor									\label{TPPareto_eliminate2}		
			\State \Return $(\Psi_{TP} - nPList)$	\label{TPPareto_return}
		\EndFunction		
	\end{algorithmic}
\end{algorithm}

\section{Hybrid Heterogeneous Server Energy Modeling} \label{sec:Energy Modeling}

We describe our solution method here to solve the problem of modelling the dynamic energy consumption during application execution on a hybrid server composed of heterogeneous computing elements. The method is based purely on system level measurements.

To motivate the case for modelling, let us consider the optimization problem for minimizing the dynamic energy consumption during the parallel execution of a workload. To obtain the optimal workload distribution, a na\"ive approach explores all possible workload distributions. For each workload distribution, it determines the total dynamic energy consumption during the parallel execution of the workload from the system-level energy measurement. It returns the workload distribution with the minimum total dynamic energy consumption. This approach, however, has exponential complexity.

Therefore, to reduce this complexity, we need energy models of the heterogeneous computing elements that can then be input to \emph{HEPOPTA} to determine the workload distribution minimizing the dynamic energy consumption during the parallel execution of the workload.

Our solution method comprises of two main steps. The first step is the identification or grouping of the computing elements satisfying properties that allow measurement of their energy consumptions to sufficient accuracy. We call these groups as \textit{abstract processors}. The second step is the construction of the dynamic energy models of the abstract processors where the principal goal apart from minimizing the time taken for model construction is to maximize the accuracy of measurements.

\subsection{Grouping of Computing Elements} \label{abstract_processors_formulation}

We group individual computing elements executing an application together in such a way that we can accurately measure the energy consumption of the group. We call these groups \textit{abstract processors}. We consider two properties essential to composing the groups:
\begin{itemize}
\item \emph{Completeness:} An abstract processor must contain only those computing elements which execute the given application kernel.
\item \emph{Loose coupling:} Abstract processors do not interfere with each other during the application. That is, the dynamic energy consumption of one abstract processor is not affected by the activities of other abstract processor.
\end{itemize}

Based on this grouping into abstract processors, we hypothesize that the total dynamic energy consumption during an application execution will equal the sum of energies consumed by all the abstract processors. So, if $E_T$ is the total dynamic energy consumption of the system incorporating $p$ abstract processors $\{AP_1,\cdots,AP_p\}$, then

\begin{equation} \label{eq:Abstract processor total energy}
E_T = \sum_{i=1}^{p} E_T(AP_i)
\end{equation}

where $E_T(AP_i)$ is the dynamic energy consumption of the abstract processor $AP_i$. We call this our \emph{additive} hypothesis.

\subsection{Energy Models of Abstract Processors} \label{abstract_processors_energy_modelling_methodology}

We describe here the second main step of our solution method, which is to build the dynamic energy models of the $p$ abstract processors. We represent the dynamic energy model of an abstract processor by a discrete function composed of a set of points of cardinality $m$.

The total number of experiments available to build the dynamic energy models is $(2^p -1) \times m$. Consider, for example, three abstract processors $\{A,B,C\}$. The experiments can be classified into following categories: $\{A,B,C,\{AB,C\},\{A,BC\},\{AC,B\},ABC\}$. The category $\{AB,C\}$ represents parallel execution of application kernels on $A$ and $B$ followed by application kernel execution on $C$. For each workload size $x$, the total dynamic energy consumption is obtained from the system-level measurement for this combined execution of kernels. The categories $\{AB,C\}$ and $\{BA,C\}$ are considered indistinguishable. There are $m$ experiments in each category. The goal is to construct the dynamic energy models of the three abstract processors $\{A,B,C\}$ from the experimental points to sufficient accuracy. 

We reduce the number of experiments to $p \times m$ by employing our additive hypothesis.

\section{Experimental Results}	\label{sec:exprimental-results}

We first study the \emph{additive} approach for determining dynamic energy functions using the two data-parallel applications, matrix multiplication and 2D-FFT, on the platform consisting of two connected heterogeneous multi-accelerator NUMA nodes, \emph{HCLServer01} and \emph{HCLServer02}. We then experimentally analyse the practical performance of \emph{HEPOPTA} and \emph{HTPOPTA} on the same platform.

\subsection{Construction of Dynamic Energy Functions}

Based on our additive approach, we group the processing units of the platform into five abstract processors following the properties explained in section \ref{abstract_processors_formulation}. We name the abstract processors on \emph{HCLServer01} as CPU\_1, GPU\_1, Phi\_1, and on \emph{HCLServer02}, as CPU\_2 and GPU\_2.

The execution time and the dynamic energy functions of the abstract processors are experimentally built separately using an automated build procedure using five parallel processes where one process is mapped to one abstract processor. To ensure the reliability of our experimental results, we follow a detailed statistical methodology explained in Section \ref{sec:append}. Briefly, to obtain a data point for each function, the software uses Student's t-test and executes the application repeatedly until the sample mean of the measurement (execution time\textbackslash dynamic energy\textbackslash total energy) lies in the user-defined confidence interval and a user-defined precision is achieved. We set the confidence interval as $95\%$ and the precision as $10\%$ for our experiments.

We use an automated tool \emph{HCLWATTSUP} \cite{HCLWattsUp} to determine the dynamic energy and total energy consumptions of a given application kernel. \emph{HCLWATTSUP} has no extra overhead and therefore does not influence the energy consumption of the application kernel. We explain \emph{HCLWATTSUP} in Section \ref{sec:append}.

To eliminate the potential disturbance due to components such as SSD (Solid State Drives) and fans, we take several precautions in computing energy measurements. We explain all these measures and precautions in Section \ref{sec:append}. 

We measure the execution time of all the abstract processors executing the same workload simultaneously, thereby taking into account the influence of resource contention. The execution time for accelerators includes the time taken to transfer data between the host and devices. Figures \ref{fig:dgemmfullspeed} and \ref{fig:fftfullspeed} show the speed functions of abstract processors for matrix multiplication and 2D-FFT applications.

We build dynamic energy functions for each abstract processor using the methodology explained in section \ref{abstract_processors_energy_modelling_methodology}. To verify if \emph{additive} hypothesis is valid, we build four profiles for \emph{HCLServer01} (one parallel and one for each of the three abstract processors), and three profiles for \emph{HCLServer02} (one parallel and one for each of the two abstract processors).

\begin{figure}[!t]
	\centering
	\includegraphics[width=3.5in]{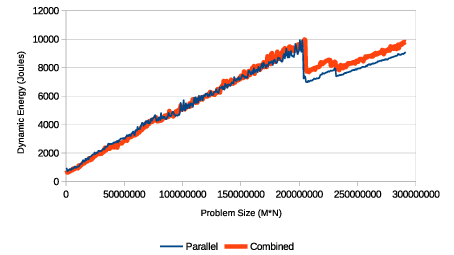}
	\caption{Parallel and Combined dynamic energy profiles for matrix multiplication application. Each data point shows the speed of execution of a problem size $M \times N$, where $M$ ranges from $64$ to $28800$ and $N$ is $10112$.}
	\label{fig:parallel_combined_DGEMM}
\end{figure}

\begin{figure}[!t]
	\centering
	\includegraphics[width=3.5in]{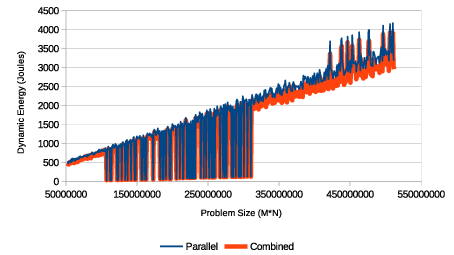}
	\caption{Parallel and Combined dynamic energy profiles for 2D-FFT application. The application calculates the 2D-DFT of a matrix size $M \times N$, where $M$ ranges from $1024$ to $10000$ and $N$ is $51200$.}
	\label{fig:parallel_combined_FFT}
\end{figure}

Figures \ref{fig:parallel_combined_DGEMM} and \ref{fig:parallel_combined_FFT} show the parallel and combined dynamic energy profiles of matrix multiplication and FFT. Here, combined refers to the sum of dynamic energy consumption of all abstract processors when running the given workload sequentially. Figure \ref{fig:dgemmfullenergy} illustrates the individual dynamic energy profiles of matrix multiplication and figure \ref{fig:fftenergy_wPhi} shows the individual dynamic energy profiles for 2D-FFT for each abstract processor. Table \ref{table:parallel_vs_combined} shows the statistics for percentage difference of parallel to combined.

We find an average difference of $5.9\%$ and $8.3\%$ between parallel and combined dynamic energy profiles on both \emph{HCLServer01} and \emph{HCLServer02} for matrix multiplication and 2D-FFT. Despite the percentage error, both parallel and combined profiles follow the same pattern for both applications. 

The parallel profiles always consume more energy than the combined profiles due to two reasons: a). Resource contention and NUMA when all abstract processors execute the given workload in parallel (which are not present when executed sequentially). This can be seen from the relatively higher error rate for \emph{HCLServer01} compared to \emph{HCLServer02} since \emph{HCLServer01} contains three abstract processors whereas \emph{HCLServer02} has two abstract processors. b). The high precision setting of 10\% for our experiments, which means that \emph{HCLWATTSUP} keeps executing the given application workload until the sample mean lies in the precision interval of 10\%. 

We can reduce the error rate between parallel and combined dynamic energy consumption significantly if we set the precision to 2.5\%. It will however drastically increase the execution time to determine the sample mean for the given experimental data point since we have to build seven profiles: four on \emph{HCLServer01} and three on \emph{HCLServer02}. We will study in our future work how to leverage the \emph{additive} component energy modelling without incurring a significant time penalty.

\begin{table}
	\caption{Percentage difference of dynamic energy consumption of parallel to combined.}
	\label{table:parallel_vs_combined}
	\centering
	\begin{tabular}{ |l|l|l|l|l| }
		\hline
		\textbf{Platform} & \textbf{Application} & \textbf{Min} & \textbf{Max} & \textbf{Average} \\ \hline
		HCLServer01 & DGEMM & 0.026\% & 29.2\% & 6.38\% \\ \hline
		HCLServer02 & DGEMM &  0.001\% & 29.03\% & 3.8\% \\ \hline
		BOTH & DGEMM &  0.04\% & 26.1\% & 5.9\% \\ \hline
		\\ \hline
		HCLServer01 & 2D-FFT & 1.8\% & 18.4\% & 9.1\% \\ \hline
		HCLServer02 & 2D-FFT &  0.02\% & 28.8\% & 12.4\% \\ \hline
		BOTH & 2D-FFT &  0.16\% & 24.7\% & 8.3\% \\ \hline
	\end{tabular}
\end{table}

\subsection{Analysis of \emph{HEPOPTA}} \label{exp:analysis:hepopta}

The experimental data set for matrix multiplication is $\{64\times10112, 128\times10112, 196\times10112, \cdots, 57600\times10112\}$, and for 2D-FFT is $\{1024\times51200, 1040\times51200, 1056\times51200, \cdots, 20000\times51200\}$. We determine the minimum, average and maximum cardinality of globally Pareto-optimal sets determined by \emph{HEPOPTA}. These values for the matrix multiplication application are ($1$, $55$, $96$), and for the 2D-FFT application, ($1$, $11$, $33$). 

We study improvements in performance and reductions in dynamic energy consumption of optimal solutions determined by \emph{HEPOPTA} in comparison with load balanced workload distribution. A load balance distribution is one with the minimum difference between the execution times of processors. The number of active processors in load-balanced solutions may be less than the total number of processors. The percentage performance improvement is obtained using $(t_{balance}-t_{opt})/t_{opt}*100$, where $t_{balance}$ represents the execution time of the load balance distribution, and $t_{opt}$ is the optimal execution time. The percentage dynamic energy saving is calculated as $(e_{balance}-e_{opt})/e_{opt}*100$, where $e_{balance}$ represent the dynamic energy consumption of load balance distribution, and $e_{opt}$ is optimal dynamic energy consumption. For matrix multiplication, the average and maximum performance improvements are $26\%$ and $102\%$. The average and maximum energy saving are $130\%$ and $257\%$. For 2D-FFT, the average and maximum performance improvements are $7\%$ and $44\%$. The average and maximum dynamic energy savings are found to be $44\%$ and $105\%$.

We obtain to what extent performance can be improved when the dynamic energy consumption is increased by up to $5\%$ over the optimal and to what extent dynamic energy can be reduced with 5\% degradation in performance over the optimal. The percentage performance improvement is obtained using $(t_{e_{opt}}-t_{e_{opt} \times 1.05})/{t_{e_{opt} \times 1.05}} * 100$, where $t_{e_{opt}}$ and $t_{e_{opt} \times 1.05}$ are the execution time of the energy-optimal endpoint and execution time associated with $5\%$ increase in energy consumption over the optimal. The percentage dynamic energy saving is obtained using $(e_{t_{opt}} - e_{t_{opt} \times 1.05})/{e_{t_{opt} \times 1.05}} * 100$, where $e_{t_{opt}}$ and $e_{t_{opt} \times 1.05}$ are the dynamic energy consumption of the performance-optimal endpoint in the Pareto-optimal front and the dynamic energy consumption associated with $5\%$ degradation in performance over the optimal.

The average and maximum performance improvements for the matrix multiplication application are $5\%$ and $50\%$. These values for the 2D-FFT application are $19\%$ and $109\%$. The average and maximum savings of dynamic energy consumption for our matrix multiplication application are $18\%$ and $116\%$, and for the 2D-FFT are $6\%$ and $63\%$.

\subsection{Analysis of \emph{HTPOPTA}}

We use the same experimental data sets as those employed for analysis of \emph{HEPOPTA}. 

\begin{figure}[!t]
	\centering
	\includegraphics[width=3.5in]{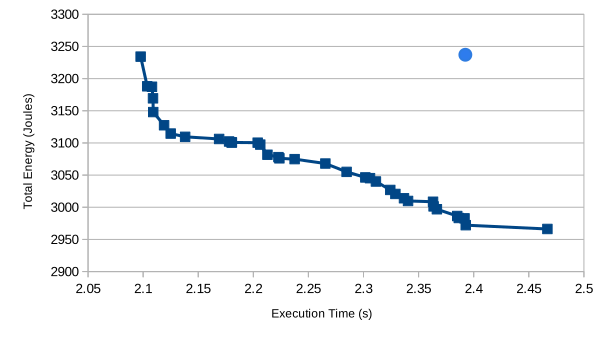}
	\caption{Globally Pareto-front solutions of execution time and total energy for a workload size $41728 \times 10112$ with the maximum cardinality determined by \emph{HTPOPTA} for the matrix multiplication application. The blue circle represents the load-balanced solution.}
	\label{fig:dgemm_TT_pareto_max}
\end{figure}

\begin{figure}[!t]
	\centering
	\includegraphics[width=3.5in]{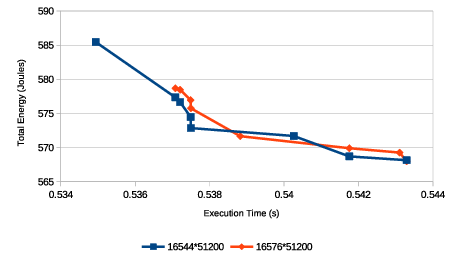}
	\caption{Globally Pareto-front solutions of execution time and total energy with the maximum cardinality determined by \emph{HTPOPTA} for the 2D-FFT application. Each curve represents a workload size.}
	\label{fig:fft_TT_pareto_max}
\end{figure}

First, the minimum, average and maximum cardinality of globally Pareto-optimal sets for execution time and total energy are determined. These values for the matrix multiplication application are $(1, 15, 35)$, and for the 2D-FFT application are $(1, 2, 8)$. The cardinalities are less than the corresponding values for the globally Pareto-optimal sets for execution time and dynamic energy since the Pareto-optimal set for execution time and total energy is a subset of Pareto-optimal set for execution time and dynamic energy. Globally Pareto-optimal sets of execution time and total energy with the maximum cardinality for matrix multiplication and FFT are shown in Figures \ref{fig:dgemm_TT_pareto_max} and \ref{fig:fft_TT_pareto_max}. In Figure \ref{fig:dgemm_TT_pareto_max}, the point above the Pareto-optimal solutions represents the execution time and total energy consumption of load-balanced distribution. The load-balanced solutions in Figure \ref{fig:fft_TT_pareto_max} have not been shown because of being far away from the sets.

We study the trade-off between execution time and total energy consumption. We calculate how much performance can be gained in case the total energy consumption is increased by up to $5\%$ over the optimal and to what extent dynamic energy can be reduced with 5\% degradation in performance over the optimal. The percentage of performance improvement is calculated using the formula $(t_{te_{opt}}-t_{te_{opt} \times 1.05}){t_{te_{opt} \times 1.05}}*100$, where $t_{te_{opt}}$ and $t_{te_{opt} \times 1.05}$ are the execution time of the total energy-optimal endpoint and execution time associated with $5\%$ increase in total energy consumption over the optimal. The percentage energy saving is obtained using the formula $(te_{t_{opt}}-te_{t_{opt} \times 1.05})/{te_{t_{opt} \times 1.05}}*100$, where $te_{t_{opt}}$ and $te_{t_{opt} \times 1.05}$ are the total energy consumption of the performance-optimal endpoint in the Pareto-optimal front and the total energy consumption associated with $5\%$ degradation in performance over the optimal.

The average and maximum performance improvements for the matrix multiplication application are $8\%$ and $17\%$. These values for the 2D-FFT application are $0.7\%$ and $9\%$. The average and maximum savings of total energy consumption for the matrix multiplication application are $4\%$ and $13\%$, and for the 2D-FFT are $0.4\%$ and $6\%$.

Using \emph{HEPOPTA}, one can find workload distributions minimizing dynamic energy consumption. \emph{HTPOPTA} provides workload distributions which minimize total energy consumption. To demonstrate that dynamic energy optimization does not always result in minimizing total energy, we calculate the percentage total energy saving over \emph{HEPOPTA} solutions for the aforementioned data set. Total energy saving is calculated using the formula, $(te_{e_{opt}}-te_{opt})/{te_{opt}}*100$, where $te_{e_{opt}}$ is total energy consumption of the solution with optimal dynamic energy consumption and $te_{opt}$ is the optimal total energy consumption. Zero percentage total energy saving represents that the same workload distribution is determined by \emph{HTPOPTA} and \emph{HEPOPTA}. The minimum, average and maximum total energy savings for the matrix multiplication application are $0$, $11\%$ and $37\%$. These values for the 2D-FFT application are $0$, $29\%$ and $106\%$.

\section{Conclusion} \label{sec:conclusion}

Performance and energy are the two most important objectives for optimization on modern parallel platforms such as supercomputers, heterogeneous HPC clusters, and cloud computing infrastructures. Recent research \cite{LastovetskyReddy2017,manumachu2018bi,manumachu2018bicpe} demonstrated the importance of workload distribution as a key decision variable in the bi-objective optimization of data-parallel applications for performance and energy on homogeneous multicore CPU clusters.

We discovered in this work that moving from single objective optimization for performance or energy to their bi-objective optimization on heterogeneous processors results in a drastic increase in the number of optimal solutions (workload distributions) even in the simple case of linear performance and energy profiles. Motivated by this finding, we studied the full performance and dynamic energy profiles of two data-parallel applications executed on two connected heterogeneous multi-accelerator NUMA nodes and found them to be non-linear and complex, and therefore difficult to approximate as analytical functions that can be used as inputs to exact mathematical algorithms or optimization softwares for determining the globally Pareto-optimal front. 

We then proposed efficient global optimization algorithms solving the bi-objective optimization problems on heterogeneous HPC platforms for performance and dynamic energy and for performance and total energy. The decision variable, which is the \emph{workload distribution}, is the same for both the optimization problems. The algorithms take as input discrete speed and dynamic energy functions (for any arbitrary shape) and return the globally Pareto-optimal set of solutions (generally speaking load imbalanced). Since the algorithms required accurate dynamic energy functions as input, we presented a novel methodology addressing a fundamental challenge, which is to accurately model the energy consumption of a hybrid scientific application kernel executing on a heterogeneous HPC platform incorporating different computing devices such as a multicore-CPU, GPU, and a Xeon PHI. The methodology is purely based on system-level energy measurements.

We experimentally analysed our algorithms using two data-parallel applications, matrix multiplication and 2D fast Fourier transform. We demonstrated that solutions provided by our algorithms significantly improve the performance and reduce the energy consumption in comparison with the load-balanced configuration of the applications. We have shown that our algorithms determine a superior Pareto-optimal front containing all \emph{strong} load imbalanced solutions that are totally ignored by \emph{load balancing} approaches and best load balanced solutions.

\section{Appendices}	\label{sec:append}

The supporting materials for the main manuscript are:

\begin{itemize}
	\item Apart from the total energy, the rationale behind considering dynamic energy consumption in our problem formulations, energy modelling, and algorithms.
	\item Studying trade-off solutions for linear speed and energy functions.	
	\item Experimental methodology obtained to construct a data point in the discrete speed and energy functions.
	\item The formal description of the algorithm, \emph{HEPOPTA}, and the helper routines used in it.
	\item Correctness and complexity proofs of \emph{HEPOPTA}.
	\item Formal description of \emph{HTPOPTA} and its correctness and complexity proofs.
\end{itemize}

\subsection{Static and Dynamic Energy Consumptions}

There are two types of energy consumptions, static energy, and dynamic energy. We define the static energy consumption as the energy consumption of the platform without the given application execution. Dynamic energy consumption is calculated by subtracting this static energy consumption from the total energy consumption of the platform during the given application execution.  The static energy consumption is calculated by multiplying the idle power of the platform (without application execution) with the execution time of the application. That is, if $P_S$ is the static power consumption of the platform, $E_T$ is the total energy consumption of the platform during the execution of an application, which takes $T_E$ seconds, then the dynamic energy $E_D$ can be calculated as,

\begin{equation} \label{eq:DynamicEnergy}
\begin{aligned}
E_D = E_T - (P_S \times T_E)
\end{aligned}
\end{equation}

The rationale behind attaching importance to dynamic energy consumption and excluding static energy consumption is the following:
\begin{enumerate}
	\item Static energy consumption is a hard constant (or an inherent property) of a platform that can not be optimized. That is, it does not depend on the application configuration and will be the same for different application configurations.
	\item Although static energy consumption is a major concern in embedded systems, it is becoming less compared to the dynamic energy consumption due to advancements in hardware architecture design in HPC systems. 
	\item We target applications and platforms where dynamic energy consumption is the dominating source of energy dissipation.
	\item Finally, we believe the inclusion of static energy consumption can underestimate the true worth of our optimization technique or any optimization technique that minimizes the dynamic energy consumption. For example, let us consider two platforms. The first platform contains nodes with just multicore CPUs. The second platform contains nodes, where each node has similar multicore CPUs that are connected to many accelerators via PCI-E links (plus multiple hard disks and fans). If we include static energy consumption, the results demonstrated by an energy prediction model on the second platform will be far inferior compared to the first platform. This is because the static energy consumption of multicores plus accelerators (plus PCI-E links, hard disks, fans, etc) will dominate the total energy consumption in the case of the second platform.
\end{enumerate}

\subsection{Solving the \emph{BOPPE} With Linear Execution Time and Dynamic Energy Functions}

\begin{proposition}		\label{prop:linear2}
	Suppose there are two processors $P_0$ and $P_1$ with linear time functions, $t_i(x)=a_i \times x$, and linear dynamic energy functions of problem size, $e_i(x) = b_i \times x$, where $i \in \{0,1\}, a_i \in \mathbb{R}_{>0}, b_i \in \mathbb{R}_{>0}$. For any given workload size $n$, the Pareto-front of solutions for execution time and dynamic energy will be linear. The decision variable is the workload distribution.
\end{proposition}

\begin{figure}[!t]
	\centering
	\includegraphics[width=3.5in]{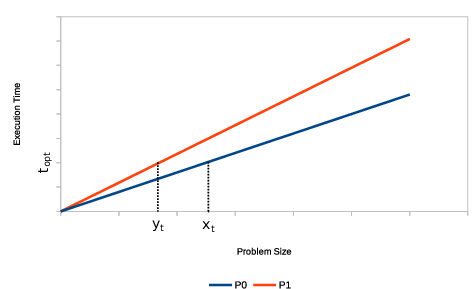}
	\caption{The Linear time functions for given processors $P_0$ and $P_1$.}
	\label{fig:proof_time}
\end{figure}

\textit{Proof.} Suppose there exist two linear time functions $t_0(x) = a_0 \times x$ and $t_1(x) = a_1 \times x$ as shown in Figure \ref{fig:proof_time}. We assume that $0 < a_0 < a_1$.

\begin{figure}[!t]
	\centering
	\includegraphics[width=3.5in]{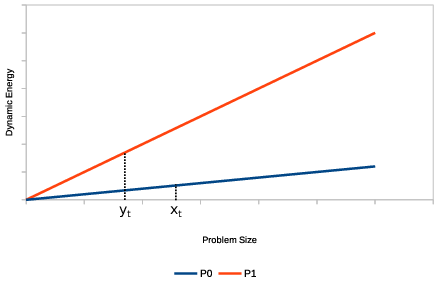}
	\caption{The Linear dynamic energy functions for given processors $P_0$ and $P_1$.}
	\label{fig:proof_energy}
\end{figure}

The linear dynamic energy profiles $e_0(x) = b_0 \times x$, and $e_1(x) = b_1 \times x$ for $P_0$ and $P_1$ are shown in Figure \ref{fig:proof_energy}. It is assumed that $0 < b_0 < b_1$.

Consider $(x_t,y_t)$ to be a workload distribution with the minimum execution time ($t_{opt}$) for executing $n$ on the two processors. $x_t$ and $y_t$ represents the problem sizes given to $P_0$ and $P_1$ where $x_t+y_t=n$. Since functions are linear, the workload distribution $(x_t,y_t)$ balances the load between processors, that is $t_0(x_t) = t_1(y_t) \implies a_1 \times x_t = a_2 \times y_t$. The dynamic energy consumption of the distribution is equal to $e_{t_{opt}} = b_0 \times x_t + B_1 \times y_t$. The distribution is shown in Figures \ref{fig:proof_time} and \ref{fig:proof_energy}. 

All possible workload distributions $(x,y)$ for the problem size $n$ are formulated below.

\begin{enumerate}
	\item $S_1 = \{(x,y) | x = x_t + \Delta, y = y_t - \Delta, \Delta \in [0,y_t]\}$
	\item $S_2 = \{(x,y) | x = x_t - \Delta, y = y_t + \Delta, \Delta \in (0,x_t]\}$
\end{enumerate}

To prove which workload distribution $(x,y)$ involves in the Pareto-front set, we study the execution time and dynamic energy consumption of all workload distributions in the both sets $S_1$ and $S_2$.

\begin{itemize}
	\item \textbf{Set $S_1$:} We prove that execution time constantly increase and dynamic energy steadily decrease as $\Delta$ grows. The execution time of a distribution $(x,y)$ can be calculated as: $t(x_t + \Delta + y_t - \Delta) = \max(t_0(x_t + \Delta), t_1(y_t - \Delta)) = \max (a_0 \times x_t + a_0 \times \Delta, a_1 \times y_t - a_1 \times \Delta) = a_0 \times x_t + a_0 \times \Delta$ $\implies$ $t(x_t + \Delta + y_t - \Delta) = a_0 \times x_t + a_0 \times \Delta$. Thus, it can be concluded that all execution times are greater than $t_{opt} = \max(a_0 \times x_t , a_1 \times y_t)$. Since the first derivative of the functions, $\frac{d t(x_t + \Delta + y_t - \Delta)}{d \Delta}$, is equal to the positive constant value $a_0$, \textit{the execution times constantly increase by growing $\Delta$}.
	
	The dynamic energy consumption of a distribution $(x,y)$ is obtained as: $e(x_t + \Delta + y_t - \Delta) = e_0(x_t + \Delta) + e_1(y_t - \Delta) = b_0 \times x_t + b_0 \times \Delta + b_1 \times y_t - b_1 \times \Delta = b_0 \times x_t + b_1 \times y_t + (b_0 - b_1) \Delta$. Due to $0 < b_0 < b_1$, it can be concluded that all dynamic energies are less than $e_{t_{opt}} = b_0 \times x_t + b_1 \times y_t$. Because the first derivative of the energy function, $\frac{d e(x_t + \Delta + y_t - \Delta)}{d \Delta}$ equals the negative constant $b_0 - b_1$, \textit{the consumed dynamic energy steadily decreases as $\Delta$ is growing}. 
	\item \textbf{Set $S_2$:} It is proved that execution time and dynamic energy are both steadily increase by growing $\Delta$. The execution time of a distribution $(x,y)$ can be calculated as: $t(x_t - \Delta + y_t + \Delta) = \max (a_0 \times x_t - a_0 \times \Delta, a_1 \times y_t + a_1 \times \Delta) = a_1 \times y_t + a_1 \times \Delta$ $\implies$ $t(x_t - \Delta + y_t + \Delta) = a_1 \times y_t + a_1 \times \Delta$. Thus, one can conclude that all execution times are greater than $t_{opt} = \max(a_0 \times x_t , a_1 \times y_t)$. Since the first derivative of the functions, $\frac{d t(x_t + \Delta + y_t - \Delta)}{d \Delta}$, is a positive constant value ($a_1$), \textit{the execution times constantly increase by growing $\Delta$}.
	
	The dynamic energy consumption of a distribution $(x,y)$ is obtained as: $e(x_t - \Delta + y_t + \Delta) = b_0 \times x_t - b_0 \times \Delta + b_1 \times y_t + b_1 \times \Delta = b_0 \times x_t + b_1 \times y_t + (b_1 - b_0) \Delta$. Because it is assumed that $0 < b_0 < b_1$, one can be concluded that all dynamic energies are greater than $e_{t_{opt}} = b_0 \times x_t + b_1 \times y_t$. Since the first derivative of the energy function, $\frac{d e(x_t + \Delta + y_t - \Delta)}{d \Delta}$ is a positive constant $b_1 - b_0$, \textit{the consumed dynamic energy steadily increase as $\Delta$ is growing}. 
\end{itemize}

Going by the definition of Pareto-optimality, all workload distributions $(x,y)$ in the set $S_1$ are Pareto-optimal solutions. No distribution, however, in $S_2$ is a Pareto-optimal solution. Henceforth, according to distributions in $S_1$, the Pareto-optimal set for for the problem size $n$ can be formulated as: $\{(t(\Delta),e(\Delta) | t(\Delta) = a_0 \times x_t + a_0 \times \Delta, e(\Delta) = b_0 \times x_t + b_1 \times y_t + (b_0 - b_1) \times \Delta, \Delta \in [0,y_t]\}$. 

Now, we will prove that all these points fall on a straight line by showing that $e(\Delta)$ is a linear function of $t(\Delta)$. We know $t(\Delta) = a_0 \times x_t + a_0 \times \Delta$. Then, $\Delta$ can be obtained as a function of $t(\Delta)$:
\begin{equation}	\label{eq:delta}
\Delta = \frac{t(\Delta) - a_0 \times x_t}{a_0} 
\end{equation}

We replace $\Delta$ in $e(\Delta)$ with Equation \ref{eq:delta}: $e(\Delta) = b_0 \times x_t + b_1 \times y_t + (b_0 - b_1) \times \frac{t(\Delta) - a_0 \times x_t}{a_0}$. Since $a_0$, $b_0$, $b_1$, $x_t$ and $y_t$ are constant, $e(\Delta)$ can be simplified as: $e(\Delta) = z_1 \times t(\Delta) + z_0$, where $z_0$ and $z_1$ are constant, which determines a linear relationship between $t(\Delta)$ and $e(\Delta)$.

To summarise, we classify all possible workload distributions for a given problem size $n$ into two groups: $S_1$ and $S_2$. We prove that no distributions in $S_2$ result in Pareto-optimal solution. However, all distributions in $S_1$ compose a linear Pareto-optimal front consisting of infinite solutions. Because every solution in $S_1$ has its unique execution time and dynamic energy consumption, one can conclude that there is a one-to-one mapping between the Pareto-optimal solutions and the workload distributions in $S_1$.

In the same manner, we can prove the correctness of the Proposition when $0 < a_0 < a_1$ and $0 < b_1 < b_0$. 
\textit{End of Proof}.

\begin{proposition}	
	Suppose there are an arbitrary number of processors ($p \ge 2$) with linear execution time and dynamic energy functions. For any given workload size, the Pareto-front for execution time and dynamic energy will contain an infinite number of solutions. The decision variable is the workload distribution.
\end{proposition}

\textit{Proof.} We prove this proposition using mathematical induction.

\begin{enumerate}
	\item Regarding Proposition \ref{prop:linear2}, solving \emph{BOPPE} for a given workload size $n$, results in an infinite number of solutions for $p = 2$ processors with linear execution time and dynamic energy functions.
	\item Assume the proposition is true for $p = k$, $k \in \{3, 4, 5, \cdots\}$.
	\item Prove that there is an infinite number of Pareto-optimal solutions for the workload size $n$ executing on $p = k+1$ processors with linear execution time and dynamic energy functions.
\end{enumerate}

Consider a problem size $\epsilon << n$ given to the fastest processor where its execution time ($t(\epsilon)$) is less than that of the other $k$ processors, executing the workload size $n - \epsilon$. According to the induction assumption, solving \emph{BOPPE} for $n - \epsilon$ on $k$ processors leads to an infinite number of solutions. Since it is supposed that $t(\epsilon)$ is the smallest execution time, the Pareto-optimal set for the workload $n$ on $k + 1$ processors is the same as the set for $n - \epsilon$ on the $k$ processors, with an infinite cardinality.

Therefore, the proposition is proved to be true for all $p \ge 2$. \textit{End of Proof}.

\subsection{Experimental Methodology to Obtain a Data Point}

To make sure the experimental results are reliable, we follow the methodology described below:

\begin{itemize}
	\item The server is fully reserved and dedicated to these experiments during their execution. We also ensure that there are no drastic fluctuations in the load due to abnormal events in the server by monitoring its load continuously for a week using the tool \textit{sar}. Insignificant variation in the load was observed during this monitoring period suggesting normal and clean behaviour of the server.
	\item Our hybrid application is executed simultaneously on all the three abstract processors, CPU, GPU, and Xeon Phi. To obtain a data point in both speed and dynamic energy functions, the application is repeatedly executed until the sample mean lies in the 95\% confidence interval and a precision of 0.1 (10\%) has been achieved. For this purpose, Student's t-test is used assuming that the individual observations are independent and their population follows the normal distribution. We verify the validity of these assumptions by plotting the distributions of observations.
	\item We set \emph{OMP\_PLACES} and \emph{OMP\_PROC\_BIND} environment variables to bind all the threads of a hybrid application to CPU cores.	
\end{itemize}

\subsection{Methodology to Measure Execution Time and Energy Consumption}	\label{sec:measure_method}

Suppose there exists a hybrid application, which is named \emph{app}, consisting of three sample kernels, \emph{Kernel\_{cpu}}, \emph{Kernel\_{gpu}} and \emph{Kernel\_{phi}}, which run in parallel. The goal is to measure the execution time and the dynamic energy consumption of kernels in the application. To do this, we instrument the sample application as shown in Algorithm \ref{alg_instrumentation}. This instrumented application returns the execution time of each kernel and the energy consumption of all the three kernels.

\begin{algorithm}
	\scriptsize
	\caption{Instrumentation of a sample application (\emph{app}) consisting of three kernels, executing on CPU, GPU and PHI simultaneously.} \label{alg_instrumentation}
	\algblock[Name]{Begin}{End}
	\begin{algorithmic}[1]	
		\State \Call {HCL\_WATTSUP\_START}{ }	\label{app_wattsup_start}
		\State \#pragma parallel
		\Begin
		\State $te_{cpu}1$ $\gets$ $gettimeofday()$
		\State \hskip\algorithmicindent \Call {Kernel\_{cpu}}{ }
		\State $te_{cpu}2$ $\gets$ $gettimeofday()$
		\End
		\Begin
		\State $te_{gpu}1$ $\gets$ $gettimeofday()$
		\State \hskip\algorithmicindent \Call {Kernel\_{gpu}}{ }
		\State $te_{gpu}2$ $\gets$ $gettimeofday()$
		\End
		\Begin
		\State $te_{phi}1$ $\gets$ $gettimeofday()$
		\State \hskip\algorithmicindent \Call {Kernel\_{phi}}{ }
		\State $te_{phi}2$ $\gets$ $gettimeofday()$
		\End
		\State $energy_{app}$ $\gets$ \Call {HCL\_WATTSUP\_STOP}{ }		\label{app_wattsup_stop}
		\State $te_{cpu}$ $\gets$ $te_{cpu}2 - te_{cpu}1$
		\State $te_{gpu}$ $\gets$ $te_{gpu}2 - te_{gpu}1$
		\State $te_{phi}$ $\gets$ $te_{phi}2 - te_{phi}1$
		\State \Return $(te_{cpu}, te_{gpu}, te_{phi}, energy_{app})$	
	\end{algorithmic}
\end{algorithm}

\subsubsection {Methodology to Measure Execution Time}

We instrument each kernel in the hybrid application (\emph{app}) by using the member function \emph{gettimeofday()} of the Linux library \emph{sys/time.h} to measure its execution time separately. As shown in Algorithm \ref{alg_instrumentation}, the execution times are stored in variables $te_{cpu}$, $te_{gpu}$ and $te_{phi}$ and are returned at the end of the application execution.

\subsubsection {Methodology to Measure the Energy Consumption}

We have two heterogeneous hybrid nodes. Each node is facilitated with one WattsUp Pro power meter that sits between the wall A/C outlets and the input power sockets of the node. These power meters capture the total power consumption of the node. The power meters have data cables connected to one USB port of the node. One Perl script collects the data from the power meter using the serial USB interface. The execution of these scripts is non-intrusive and consumes insignificant power.

The power meters are periodically calibrated using an ANSI C12.20 revenue-grade power meter, Yokogawa WT210. The maximum sampling speed of the power meters is one sample every second. The accuracy specified in the data-sheets is $\pm{3\%}$. The minimum measurable power is $0.5$ watts. The accuracy at $0.5$ watts is $\pm{0.3}$ watts.

We use \emph{HCLWattsUp} API, which gathers the readings from the power meters to determine the average power and energy consumption during the execution of an application for the whole node. \emph{HCLWattsUp} API \cite{HCLWattsUp} also provides two macros: \emph{HCL\_WATTSUP\_START} and \emph{HCL\_WATTSUP\_STOP}. The \emph{HCL\_WATTSUP\_START} macro starts gathering power readings from the power meter using the aforementioned Perl script, whereas the \emph{HCL\_WATTSUP\_STOP} stops gathering and return the total energy as a sum of these power readings.

To measure the amount of energy consumed by the application, we invoke \emph{HCL\_WATTSUP\_START} and \emph{HCL\_WATTSUP\_STOP} macros as shown in Algorithm \ref{alg_instrumentation}.  The consumed energy is stored in the variable $energy_{app}$ and is returned at the end of the application execution.

\subsubsection{Methodology to Ensure Reliability of Experimental Results}

As explained in Section \ref{sec:measure_method}, each application is instrumented for measuring its performance and energy consumption. The measured execution times and consumed energy in each run of the application are stored in the variables $te_{cpu}$, $te_{gpu}$, $te_{phi}$, and $energy_{app}$, which are returned when the application execution finishes (Sample algorithm \ref{alg_instrumentation}). 

We keep running the application until the sample means of the measured execution times and energy consumption of the application lie within a given confidence interval, and a given precision is achieved. For this, we employ a script, which is named \Call{MeanUsingTtest}{}. Algorithm \ref{mean-t-test} presents the pseudocode of this script. It executes the application \emph{app} repeatedly until one of the following three conditions is satisfied:

\begin{enumerate}
	\item The maximum number of repetitions ($maxReps$) has been exceeded (Line \ref{mean-t-while}).
	\item The sample means of all devices (kernel execution times and the application energy consumption) fall in the confidence interval (or the precision of measurement $eps$ has been achieved) (Lines \ref{mean-t-sample1}-\ref{mean-t-sample2}).
	\item The elapsed time of the repetitions of application execution has exceeded the maximum time allowed ($maxT$ in seconds) (Lines \ref{mean-t-etime1}-\ref{mean-t-etime2}).
\end{enumerate}

\Call{MeanUsingTtest}{} returns the sample means of the execution times for each abstract processor (i.e. $time_{cpu}$, $time_{gpu}$, $time_{phi}$) and the energy consumption of all kernels (i.e. $energy$). The input parameters are minimum and maximum number of repetitions, $minReps$ and $maxReps$. These parameter values differ based on the problem size solved. For small problem sizes ($32 \leq n \leq 1024$), these values are set to $10000$ and $100000$. For medium problem sizes ($1024 < n \leq 5120$), these values are set to $100$ and $1000$. For large problem sizes ($n > 5120$), these values are set to $5$ and $50$. The values of $maxT$, $cl$, and $eps$ are set to $3600$, $0.95$, and $0.1$. If the precision of measurement is not achieved before the maximum number of repeats have been completed, we increase the number of repetitions and also the maximum elapsed time allowed. However, we observed that condition (2) is always satisfied before the other two in our experiments. 

\begin{algorithm}
\scriptsize
\caption{Script determining the mean of an experimental run using student's t-test.} \label{mean-t-test}
\begin{algorithmic}[1]
\Procedure{MeanUsingTtest}{$app,minReps,maxReps,maxT,cl,eps,$ \par
$reps\#, elapsedTime, time_{cpu}, time_{gpu}, time_{phi}, energy$}
\INPUT
\Statex The application to execute, $app$
\Statex The minimum number of repetitions, $minReps \in \mathbb Z_{> 0}$
\Statex The maximum number of repetitions, $maxReps \in \mathbb Z_{> 0}$
\Statex The maximum time allowed for the application to run, $maxT \in \mathbb R_{> 0}$
\Statex The required confidence level, $cl \in \mathbb R_{> 0}$
\Statex The required accuracy, $eps \in \mathbb R_{> 0}$
\OUTPUT
\Statex The number of experimental runs actually made, $reps\# \in \mathbb Z_{> 0}$
\Statex The elapsed time, $elapsedTime \in \mathbb R_{> 0}$
\Statex The mean execution times, $time_{cpu}, time_{gpu}, time_{phi} \in \mathbb R_{\ge 0}$
\Statex The mean consumed energy, $energy \in \mathbb R_{> 0}$
\Statex
     \State $reps \gets 0$; $stop \gets 0$; $etime \gets 0$
     \State $sum_{cpu} \gets 0$; $sum_{gpu} \gets 0$; $sum_{phi} \gets 0$; $sum_{eng} \gets 0$
     \While{($reps < maxReps$) and ($!stop$)}	\label{mean-t-while}
        \State $(t_{cpu}[reps],t_{gpu}[reps],t_{phi}[reps],eng[reps])$ $\gets$ \Call{Execute}{app} \label{mean-t-exec}
        \State $sum_{cpu} += t_{cpu}[reps]$
        \State $sum_{gpu} += t_{gpu}[reps]$
        \State $sum_{phi} += t_{phi}[reps]$ 
        \State $sum_{eng} += eng[reps]$ 
        \If{$reps > minReps$}
           \State $stop_{cpu}$ $\gets$ \Call {CalAccuracy}{$cl,reps+1,t_{cpu},eps$}	\label{mean-t-sample1}
           \State $stop_{gpu}$ $\gets$ \Call {CalAccuracy}{$cl,reps+1,t_{gpu},eps$}
           \State $stop_{phi}$ $\gets$ \Call {CalAccuracy}{$cl,reps+1,t_{phi},eps$}
           \State $stop_{eng}$ $\gets$ \Call {CalAccuracy}{$cl,reps+1,t_{eng},eps$}
           \State $stop$ $\gets$ $stop_{cpu} \wedge stop_{gpu} \wedge stop_{phi} \wedge stop_{eng}$ \label{mean-t-sample2}         
           \If{$\max\{sum_{cpu},sum_{gpu},sum_{phi}\} > maxT$}	\label{mean-t-etime1}
              \State $stop \gets 1$
           \EndIf												\label{mean-t-etime2}
        \EndIf
        \State $reps \gets reps + 1$
     \EndWhile
     \State $reps\# \gets reps$
     \State $elapsedTime \gets \max\{sum_{cpu},sum_{gpu},sum_{phi}\}$
     \State $time_{cpu} \gets \frac{sum_{cpu}}{reps}$; $time_{gpu} \gets \frac{sum_{gpu}}{reps}$; $time_{phi} \gets \frac{sum_{phi}}{reps}$
     \State $energy \gets \frac{sum_{eng}}{reps}$
     \State \Return $(reps\#, elapsedTime, time_{cpu}, time_{gpu}, time_{phi}, energy)$
\EndProcedure
\end{algorithmic}
\end{algorithm}

Algorithm \ref{alg_calAcc} shows the pseudocode of the helper functions \Call{CalAccuracy}{}, which is used by \Call{MeanUsingTtest}{}. It returns $1$ if the sample mean of a given reading lies in the $95\%$ confidence interval ($cl$) and a precision of $0.1$ ($eps = 10\%$) has been achieved. Otherwise, it returns $0$.

\begin{algorithm}
	\scriptsize
	\caption{Algorithm Calculating Accuracy} \label{alg_calAcc}
	\begin{algorithmic}[1]	
		\Function{CalAccuracy}{$cl,reps,Array,eps$}
			\State $clOut$ $\gets$ fabs(gsl\_cdf\_tdist\_Pinv($cl$, $reps-1$)) \par
		    \hskip\algorithmicindent\hskip\algorithmicindent\hskip\algorithmicindent
			$\times$ gsl\_stats\_sd($Array$, 1, $reps$) \par
			\hskip\algorithmicindent\hskip\algorithmicindent\hskip\algorithmicindent
			/ sqrt($reps$)
			\If{$clOut \times \frac{reps}{\sum_{i=0}^{reps-1} Array[i]} < eps$}
				\State \Return $1$
			\EndIf
			\State \Return $0$
		\EndFunction		
	\end{algorithmic}
\end{algorithm}

If the precision of measurement is not achieved before the maximum number of repeats have been completed, we increase the number of repetitions and also the maximum elapsed time allowed. However, we observed that condition (2) is always satisfied before the other two in our experiments.

\subsection{Precautions to Rule out Interference of Other Components in Dynamic Energy Consumption}

We take several precautions in computing energy measurements to eliminate any potential interference of the computing elements that are not part of the given abstract processor running the given application kernel. Consequently, it ensures that the dynamic energy of the given abstract processor computed in this way solely represents the dynamic energy consumed by the constituent computing elements of the very abstract processor. For this, we take following precautions:

\begin{enumerate}
	\item We group abstract processors in such a way that a given abstract processor must be constituting solely the computing elements which are involved to run a given application kernel. The application kernel will, in this way, only use the computing elements of the abstract processor executing it and do not use any other component for its execution. Hence, the dynamic energy consumption will solely reflect the work done by the computing elements of the given abstract processor executing the application kernel.
	
	Consider for example mkl-DGEMM application kernel executing on only abstract processor \emph{A} (comprises of CPU and DRAM). However, \emph{HCLWattsUp} API gives the total energy consumption of the server during the execution of an application. This includes the contribution from all components such as NIC, SSDs, fans, etc. Therefore, to rule out their contribution in dynamic energy consumption, we ensure all the components other than CPUs and DRAM are not used during the execution of an application. In this way, the dynamic energy consumption that we obtain using \emph{HCLWattsUp} API reflects only the contribution of CPUs and DRAM. For this, we follow the below steps to verify if these other components are not used:
	\begin{itemize}
		\item We monitor the disk consumption before and during the application run and ensure that there is no I/O performed by the application using tools such as \emph{sar}, \emph{iotop}, etc.
		\item we ensure that the problem size used in the execution of an application does not exceed the main memory and that swapping (paging) does not occur.
		\item We ensure that the network is not used by the application by monitoring using tools such as \emph{sar}, \emph{atop}, etc.
		\item We set the application kernel's CPU affinity mask using SCHED API's system call SCHED\_SETAFFINITY() respecting abstract processors formulation guidelines. Consider for example mkl-DGEMM application kernel executing on only abstract processor \emph{A}. To bind this application kernel, we set its CPU affinity mask to 11 physical CPU cores of Socket 1, and 11 physical CPU cores of Socket 2.
	\end{itemize}
	
	\item Fans are also a great contributor to energy consumption. On our platform 
	fans are controlled in two zones:
	a) zone 0: CPU or System fans, 
	b) zone 1: Peripheral zone fans. 
	There are 4 levels to control the speed of fans:
	\begin{itemize}
		\item Standard: BMC control of both fan zones, with CPU zone based on CPU temp (target speed 50\%) and Peripheral zone based on PCH temp (target speed 50\%)
		\item Optimal: BMC control of the CPU zone (target speed 30\%), with Peripheral zone fixed at low speed (fixed ~30\%)
		\item Heavy IO: BMC control of CPU zone (target speed 50\%), Peripheral zone fixed at 75\%
		\item Full: all fans running at 100\%
	\end{itemize}
	
	In all speed levels except the full, the speed is subject to be changed with temperature, and consequently, their energy consumption also changes with the change of their speed. Higher the temperature of CPU, for example, higher the fans speed of zone 0, and higher the energy consumption to cool down. This energy consumption to cool the server down, therefore, is not consistent and is dependent on the fans speed, and consequently can affect the dynamic energy consumption of the given application kernel. Hence, to rule out fans' contribution in dynamic energy consumption, we set the fans at full speed before launching the experiments. When set at full speed, the fans on our platform run consistently at $ {\sim}13400 $ rpm, and do not change their speed until we do so to another speed level. In this way, fans consumed same amount of power which is included in static power of the server. We monitor the temperature of server and speed of the fans (after setting it at full) with help of Intelligent Platform Management Interface (IPMI) sensors, both with and without the application run. We find no considerable difference in temperature, and find the speed of fans same in both scenarios.
\end{enumerate}

Thus, we ensure that the dynamic energy consumption obtained using \emph{HCLWattsUp}, reflects the contribution solely by the \emph{abstract processor} executing the given application kernel.

\subsection{Helper Routines Called in \emph{HEPOPTA}}

\subsubsection{Function \emph{ReadFunc}}

The input parameters to the function \Call{ReadFunc}{} are $F$, which can be a discrete time ($t_i, i \in \{0, 1, \cdots, p-1\}$) or dynamic energy function ($e_i, i \in \{0, 1, \cdots, p-1\}$), and a problem size $w$ (Algorithm \ref{alg_ReadFunc}). The function returns the execution time or the dynamic energy consumption of $w$ from the function $F$. It returns $0$ for zero problem sizes and $-1$ when there is no match for $w$ in the function. This function uses functions sorted by problem size.

\begin{algorithm}
	\scriptsize
	\caption{Algorithm Reading the Execution Time or Energy Consumption of a Given Problem Size} \label{alg_ReadFunc}
	\begin{algorithmic}[1]	
		\Function{ReadFunc}{$F,w$}
			\If{$w = 0$}
				\State \Return $0$
			\EndIf
			\If{$\nexists(w,f_{iw}) \in F$}
				\State \Return $-1$
			\EndIf
			\State \Return $f_{iw}$
		\EndFunction		
	\end{algorithmic}
\end{algorithm}

\subsubsection{Function \emph{SizeThresholdCalc}}

The Algorithm \ref{alg_sizeTh_hepopta} shows the pseudocode of the function \Call{SizeThresholdCalc}{} which calculates the size threshold array, $\sigma$. First, It determines the size threshold of $L_{p-1}$ by finding the greatest problem size in the energy function $E_{p-1}$ that its energy consumption is less than or equal to $\varepsilon$ (Line \ref{alg_sizeTh_hepopta_largestW1}). Then, it calculates $\sigma_i$, $i \in \{0,1,,\cdots,p-2\}$ where $\sigma_i$ is the summation of $\sigma_{i+1}$ with the greatest work-size in energy function $E_{i}$ that its consumed dynamic energy is less than or equal to $\varepsilon$ (Lines \ref{alg_sizeTh_hepopta_loop1}-\ref{alg_sizeTh_hepopta_loop2}).

\begin{algorithm}
	\scriptsize
	\caption{Algorithm Determining Size Thresholds} 		\label{alg_sizeTh_hepopta}
	\begin{algorithmic}[1]	
		\Function{SizeThresholdCalc}{$p, E, \varepsilon, \sigma$}		
			\State $\sigma_{p-1}$ $\gets$ $\max_{j=0}^{m-1}\{x_{(p-1)~j}~|~e_{(p-1)~j} \le \varepsilon\}$		\label{alg_sizeTh_hepopta_largestW1}
			\ForAll{$i = p-2$; $ i \ge 0$; $ i{-}{-}$}	\label{alg_sizeTh_hepopta_loop1}
				\State $\sigma_i$ $\gets$ $\sigma_{i+1}+\max_{j=0}^{m-1}\{x_{ij}~|~e_{ij} \le \varepsilon\}$
			\EndFor		\label{alg_sizeTh_hepopta_loop2}
			\State \Return $\sigma$
		\EndFunction		
	\end{algorithmic}
\end{algorithm}

The function uses dynamic energy functions sorted in non-decreasing order of dynamic energy consumption.

\subsubsection{Function \emph{Cut}}

The function \Call{Cut}{} returns \emph{TRUE} if the input workload $n$ is greater than the input size threshold $\sigma$ (Algorithm \ref{alg_cut_code_hepopta}).

\begin{algorithm}
	\scriptsize
	\caption{Algorithm Cutting Search Tree using the Size Threshold} \label{alg_cut_code_hepopta}
	\begin{algorithmic}[1]	
		\Function{Cut}{$n, \sigma$}
			\If{$n > \sigma$}
				\State \Return $TRUE$
			\EndIf
			\State\Return $FALSE$
		\EndFunction		
	\end{algorithmic}
\end{algorithm}

\subsubsection{Structure of matrix \emph{PMem} in \emph{HEPOPT}}	\label{sec:mem_struct}

We use \emph{PMem}, a two-dimensional array, to memorize Pareto-front solutions for dynamic energy and performance which have been found at levels $\{L_1,\cdots,L_{p-2}\}$ in solution trees. Consider a given memory cell $PMem[i][n]$ which saves a Pareto-optimal solution which is found for a given workload $n$ on processors $\{P_i,\cdots,P_{p-1}\}, i \in \{1, 2, \cdots, p-2\}$. The memory cell consists of a set where each element in this set is a tuple like $<eng,time, part, P\#, key>$, storing one Pareto-optimal solution. 

The field $eng$ stores the dynamic energy consumption of the Pareto-optimal solution on processors $\{P_i,\cdots,P_{p-1}\}$, $time$ is its parallel execution time on the processors, $part$ determines the problem size assigned to $P_i$ by the solution, $P\#$ represents the number of active processors in the solution, and $key$ is the dynamic energy consumption of a saved Pareto-optimal solution, provided it exists, for a node at the level $L_{i+1}$ labelled by $n-part$ where this Pareto-optimal solution is the partial solution for the node $n$. Since dynamic energy consumptions are unique in Pareto-optimal sets, we use this parameter for pointing to partial solutions. In fact, $key$ operates as a pointer to partial solutions.

Elements in Pareto-optimal sets are sorted in increasing order of dynamic energy consumption. If there exists no Pareto-optimal solution for the workload $n$ on the level $i$, its corresponding memory cell, $PMem[i][n]$, will contain one tuple that its $eng$ element is set to the constant value $\_NS$ (i.e. \emph{No\_Solution}).

\subsubsection{Function \emph{ReadParetoMem}}

Algorithm \ref{readmem_readM} illustrates the function \Call{ReadParetoMem}{}. Suppose we are going to retrieve the saved solutions for a given workload $n$ on $L_c$. First, $PMem[c][n]$ is read which involves the saved solutions for $n$ (Line \ref{readmem_readM}). If $PMem[c][n]$ is empty, that is this node has not been visited yet, and the function returns \emph{DUMMY} (Lines \ref{readmem_empty1}-\ref{readmem_empty2}). In this case, \emph{HEPOPTA\_Kernel} will continue with expanding this node.

Since solutions in memory cells are sorted in the increasing order of the dynamic energy consumption, we consider the energy consumption of the first element in each set as the best solution. According to the retrieved value for $eng$, the following cases might happen:

\begin{itemize}
	\item \textbf{NOT\_SOLUTION}: This case occurs when $eng$ is equal to $\_NS$ (there is no solution for $n$ on processor $\{P_c,\cdots,P{p-1}\}$) or the consumed dynamic energy of the saved solution is greater than $\varepsilon$ (Lines \ref{readmem_noSol_1} and \ref{readmem_noSol_2}).
	\item \textbf{SOLUTION}: This case occurs if the retrieved $eng$ is less than or equal to $\varepsilon$ (Line \ref{readmem_Sol}).
\end{itemize}

\begin{algorithm}
	\scriptsize
	\caption{Algorithm Retrieving Solution from Memory} \label{readmem_readM}
	\begin{algorithmic}[1]	
		\Function{ReadParetoMem}{$n, c, \varepsilon, Mem$}
			\State $pSet \gets PMem[c][n]$		\label{readmem_cSet}
			\If{$|pSet| = 0$}					\label{readmem_empty1}
				\State \Return $DUMMY$
			\EndIf								\label{readmem_empty2}
			\If{$pSet[0].eng = \_NS \vee pSet[0].eng > \varepsilon$}	\label{readmem_noSol_1}
				\State \Return $NOT\_SOLUTION$	
			\EndIf															\label{readmem_noSol_2}
			\State \Return $SOLUTION$						\label{readmem_Sol}
		\EndFunction		
	\end{algorithmic}
\end{algorithm}

\subsubsection{Function \emph{MakeParetoFinal}}
Algorithm \ref{alg_fin_hepopta} illustrates the function \Call{MakeParetoFinal}{} which finalizes the input memory cell $pmem$. As explained in the main manuscript, each memory cell is finalized when its corresponding node along with its all children in the tree are completely explored. If a node is expanded for which there is no Pareto-optimal solution, the node labelled as $\_NS$ by inserting a tuple with the constant value $\_NS$ in field $eng$ (Line \ref{finalize_insert}). This means that there is no Pareto-front solution for this node.

\begin{algorithm}
	\scriptsize
	\caption{Algorithm Finalizing Memory Cells} \label{alg_fin_hepopta}
	\begin{algorithmic}[1]	
		\Function{MakeParetoFinal}{$pmem$}
			\If{$|pmem| = 0$} 
				\State $pSet\gets(\_NS,0,0,0,0)$	\label{finalize_insert}	
			\EndIf
		\EndFunction		
	\end{algorithmic}
\end{algorithm}

\subsubsection{Function \emph{MergePartialParetoes}}		\label{sec:merge}

For every non-leaf node, \emph{HEPOPTA\_Kernel} invokes the function \Call{MergePartialParetoes}{} to build its Pareto-optimal solutions, using the Pareto-optimal sets of its children which are named \emph{partial solutions} for the node. The function then stores the new solutions in $PMem$. If there exist two workload distributions with equal dynamic energy consumption and execution time, \Call{MergePartialParetoes}{} selects the solution with the minimum number of active processors. The input variable $c$ indicates a level in the tree, and $partsVec$ is a list including all problem sizes allocated to $P_c$ where results in a solution. The algorithm starts with initializing $pSet$ which points to a memory cell storing Pareto-optimal solutions for a workload $n$ on $L_c$ (Lines \ref{merge_pset1}-\ref{merge_pset2}). The set $\Psi_{EP}$ will store final globally Pareto-optimal solutions for the root. The first $For$ loop iterates all problem sizes in $partsVec$ and builds new feasible solutions by merging the problem sizes in $partsVec$ with their corresponding partial Pareto-optimal solutions (Lines \ref{merge_for1_1}-\ref{merge_for1_2}). In each iteration, for a given problem size $x$, \Call{MergePartialParetoes}{} finds the partial Pareto-optimal solutions ($subPareto$) in $PMem$ (if $1 \le c < p-2$) or builds it (if $c = p-2$) (Lines \ref{merge_subPareto1}-\ref{merge_subPareto2}). The inner $For$ loop scans all Pareto-optimal solutions in $subPareto$. It merges the problem size $x$, given to $P_c$, with Pareto-solutions in $subPareto$ for processors $\{P_{c+1},\cdots,P_{p-1}\}$ (Lines \ref{merge_innerloop1}-\ref{merge_innerloop2}). For each merged solution, $pSet$ is examined to verify whether there exists a Pareto-optimal solution in the set. If it is the case, $pSet$ is updated, and all non-Pareto-optimal solutions are eliminated. Therefore, for each newly merged solution $(eng_x,time_x,x,P\#_x,key)$, the following situations may happen:

\begin{enumerate}
	\item $pSet$ is empty and the solution is inserted (Line \ref{merge_empty1}).
	\item There exists a solution in $pSet$ that its $eng$ is equal to $eng_x$. In this case the saved solution is updated if either $eng_x$ is less than $eng$ or $P\#_x$ is less than $P\#$ (Lines \ref{merge_equal1}-\ref{merge_equal2}).
	\item The $eng_x$ of the merged Pareto-optimal solution is greater than ones in the $pSet$. The solution is inserted in case its execution time $time_x$ is less than the last solution in $pSet$ (Lines \ref{merge_equal2}-\ref{merge_end2}).
	\item The $eng_x$ of the merged Pareto-optimal solution is less than ones in the $pSet$. The solution is inserted in $pSet$ after eliminating all non-Pareto-optimal solutions (Lines \ref{merge_end2}-\ref{merge_begin2}).
	\item The $eng_x$ of the merged solution is somewhere at the middle of $pSet$. In this case, the solution is inserted in $pSet$, and all non-Pareto-optimal solutions are removed (Lines \ref{merge_begin2}-\ref{merge_middle2}).
\end{enumerate}

It should be mentioned that the function \emph{lower\_bound} returns a pointer to the first element in the $pSet$ that its $eng$ is greater than or equal to $eng_x$.

The algorithm prevents further iteration in $pSet$ if the execution time of the last-evaluated partial Pareto-optimal solution is less than or equal to the execution time of problem size $x$ on $P_c$. In fact, the further scanning of the $pSet$ will not lead to a Pareto-optimal solution. This is because all Pareto-optimal solutions are sorted in increasing order of dynamic energy consumption, that consequently implies that the execution times are decreasing in each set. Thus, all solutions built using the following elements in the $pSet$ will have the same execution time as the execution time of the workload $x$ on $P_c$ but with greater energy consumption.

Finally, the function \Call{BuildParetoSols}{} is called to obtain the workload distribution for each solution in $\Psi_{EP}$ (Lines \ref{merge_buildsol1}-\ref{merge_buildsol2}).

\begin{algorithm}
	\scriptsize
	\caption{Algorithm Merging Partial-Pareto Solutions} \label{hepopt_merge}
	\begin{algorithmic}[1]
		\Function{\textbf{MergePartialParetoes}}{$n, p, c, E, T,partsVec,PMem,\Psi_{EP}$}
			\If{$c = 0$}							\label{merge_pset1}
				\State $pSet \gets PMem[0][0]$
			\Else
				\State $pSet \gets PMem[c][n]$
			\EndIf									\label{merge_pset2}
			\ForAll{$x \in partsVec$}				\label{merge_for1_1}
				\If{$c < p - 2$}					\label{merge_subPareto1}
					\State $subPareto \gets PMem[c + 1][n - x]$
				\Else
					\State $x' \gets n-x$
					\State $P\#_{x'} \gets (x = 0~?~0~:~1)$
					\State $time_{x'} \gets \Call{ReadFunc}{T_{P-1},x'}$
					\State $eng_{x'} \gets \Call{ReadFunc}{E_{P-1},x'}$
					\State $subPareto \gets (eng_{x'}, time_{x'},x',P\#_{x'},-)$
				\EndIf								\label{merge_subPareto2}
				\State $time_x \gets \Call{ReadFunc}{T_c,x}$
				\State $P\#_x \gets (x = 0~?~0~:~1)$
				\ForAll{$tup \in subPareto$}				\label{merge_innerloop1}
					\State $eng_x \gets tup.eng + \Call{ReadFunc}{E_{c},x}$
					\State $time_x \gets Max (tup.time, time_x)$
					\State $P\#_x \gets P\#_x + P\#_{tup}$
					\State $key \gets tup.eng$
					\If{$|pSet| = 0$}			
						\State $pSet \gets (eng_x, time_x, x, P\#_x, key)$	\label{merge_empty1}
					\Else	
						\State $tup_l \gets pSet.lower\_bound(eng_x)$
						\If{$tup_l \neq pSet.end() \wedge tup_l.eng = eng_x$}	\label{merge_equal1}
							\If{$tup_l.time > time_x$}
								\State $tup_l \gets (eng_x, time_x, x, P\#_x, key)$
								\ForAll{$r \in pSet~|~r.eng > eng_x \wedge r.time \ge time_x$}
									\State $pSet \gets pSet - r$
								\EndFor
							\ElsIf{$tup_l.time = time_x \wedge P\#_x < tup_l.P\#$}
								\State $tup_l \gets (eng_x, time_x, x, P\#_x, key)$
							\EndIf
						\ElsIf{$tup_l = pSet.end()$}						\label{merge_equal2}
							\State $tup_l \gets tup_l - 1$
							\If{$tup_l.time > time_x$}
								\State $pSet \cup (eng_x, time_x, x, P\#_x, key)$
							\EndIf
						\ElsIf{$tup_l = pSet.begin()$}
							\If{$time_x \leq tup_l.time $}					\label{merge_end2}
								\ForAll{$r \in pSet~|~r.eng > eng_x \wedge r.time \ge time_x$}
									\State $pSet \gets pSet - r$
								\EndFor
							\EndIf
							\State $pSet \cup (eng_x, time_x, x, P\#_x, key)$
						\Else												\label{merge_begin2}
							\State $tup_l \gets tup_l - 1$
							\If{$tup_l.time > time_x$}
								\State $pSet \cup (eng_x, time_x, x, P\#_x, key)$
								\ForAll{$r \in pSet~|~r.eng > eng_x \wedge r.time \ge time_x$}
									\State $pSet \gets pSet - r$
								\EndFor
							\EndIf
						\EndIf												\label{merge_middle2}
						\If{$tup.time \leq time_x$}
							\State \textbf{break}							\label{merge_break}
						\EndIf
					\EndIf
				\EndFor		\label{merge_innerloop2}
			\EndFor		\label{merge_for1_2}
			\If{$c = 0$}							\label{merge_buildsol1}
				\State \Call {BuildParetoSols}{$PMem, \Psi_{EP}$}	\label{hepopta_buildsols}
			\EndIf									\label{merge_buildsol2}
		\EndFunction
	\end{algorithmic}	
\end{algorithm}

\subsubsection{Function \emph{BuildParetoSols}}
As explained in the section \ref{sec:merge}, the set $\Psi_{EP}$ holds final globally Pareto-optimal solutions for dynamic energy and performance. Each element in $\Psi_{EP}$, which represents a Pareto-optimal solution, is a triple like $(eng,time,X)$ where $eng$ determines the dynamic energy consumption of the solution, $time$ is its execution time, and $X=\{x_0, x_1, \cdots, x_{p-1}\}$ represents the workload distribution of the solution. The function \Call{BuildParetoSols}{} determines the problem sizes given to the processors $\{P_1,\cdots,P_{p-1}\}$ by any solution in $PMem[0][0]$. 

The algorithm \ref{alg_buildPareto} shows the pseudocode of \Call{BuildParetoSols}{}. The function reads the problem sizes given to the processors $\{P_1,\cdots,P_{p-1}\}$ from $PMem$. It uses the field $key$ in each saved solution to find the corresponding partial solution and eventually the problem size give to $P_{i+1}$. Since energy consumptions are unique in any set, there is only one tuple that its dynamic energy consumption is equal to $key$ in that set.

\begin{algorithm}
	\scriptsize
	\caption{Algorithm Completing Workload Distribution for $\Psi_{EP}$} \label{alg_buildPareto}
	\begin{algorithmic}[1]	
		\Function{BuildParetoSols}{$PMem, \Psi_{EP}$}
			\ForAll{$tup \in PMem[0][0]$}
				\State $sumSize \gets tup.part$
				\State $X[0] \gets tup.part$
				\State $key_{cur} \gets tup.key$
				\ForAll{$i = 1$; $ i \le p-2$; $ i{+}{+}$}
					\State $tup_{sub} \gets \{t \in PMem[i][n - sumSize]~|~t.eng = key_{cur}\}$
					\State $X[i] \gets tup_{sub}.part$
					\State $sumSize \gets sumSize + tup_{sub}.part$
					\State $key_{cur} \gets tup_{sub}.key$
				\EndFor
				\State $X[p-1] \gets n - sumSize$
				\State $\Psi_{EP} \gets \Psi_{EP} \cup (tup.eng, tup.time, X)$
			\EndFor
		\EndFunction		
	\end{algorithmic}
\end{algorithm}

\subsection{Correctness Proof of \emph{HEPOPTA}}

\begin{proposition}
	The algorithm \emph{HEPOPTA} always returns globally Pareto-optimal solutions.
\end{proposition}

\textit{Proof.} To obtain globally Pareto-optimal solutions for the dynamic energy and performance of a given workload $n$ between $p$ processors $\{P_0,\cdots,P_{p-1}\}$, we need all possible distributions for the workload. One approach is to employ the naive algorithm exploring full tree of solutions and build the globally Pareto-optimal set which suffers from exponential complexity. \emph{HEPOPTA} enhances the naive approach using the specific operation \emph{Cut} to just explore a small fraction of the full solution tree. Therefore, the correctness of \emph{HEPOPTA} will be proved if we show that there exists no subtree ignored by the operation \emph{Cut} while contains a Pareto-optimal solution.

Consider a given node which is labelled by $n$ at a level $L_i, i \in \{0, 1, \dots, p-2\}$ in a solution tree . The operation \emph{Cut} removes the subtree growing from a node in case the workload of this node exceeds its corresponding size threshold. Suppose the workload distribution $X=\{x_0,\cdots,x_{p-1}\}$ is eliminated from the search space by using \emph{Cut} operation. Regarding the definition of size thresholds, the dynamic energy consumption of this workload ($E_D(X) = \sum_{i=0}^{p-1}E_i(x_i)$) is greater than $\varepsilon$ ($\varepsilon < E_D(X)$). It should be mentioned that the execution time of this solution ($T_E(X) = \max_{i=0}^{p-1}T_i(x_i)$) will be greater than or equal with the optimal execution time for the workload $n$, $t_{opt}$ ($t_{opt} \le T_E(X)$). As explained in the main manuscript, $\varepsilon$ is set to the dynamic energy consumption of the optimal distribution for execution time. Hence, there is a distribution like $X^* = \{x_0^*\cdots,x_{p-1}^*\}$ where its execution time and dynamic energy consumption are $T_E(X^*) = t_{opt}$ and $E_D(X^*) = \varepsilon$, respectively. Thus, we have $E_D(X^*) < E_D(X)$ and $T_E(X^*) \le T_E(X)$, and according to the definition of Pareto-optimal sets, the solution $X$, which is removed by \emph{Cut}, goes after the solutions $X^*$ and cannot be a member of the Pareto-optimal set.
\textit{End of Proof}.

\subsection{Complexity of \emph{HEPOPTA}}

\begin{lemma}	\label{lemma:paretoSols}
	The maximum number of Pareto-optimal solutions for dynamic energy and performance on a heterogeneous platform including $p$ discrete dynamic energy and $p$ performance functions with a cardinality of $m$ is equal to $m \times p$.
\end{lemma}

\textit{Proof.} We know that the execution time of a Pareto-optimal solution with the workload distribution $X=\{x_0,x_1,\cdots,x_{p-1}\}$ is equal to the execution time of an $x_i \in X$ where $T_i(x_i) = \max_{j=0}^{p-1}T_j(x_j)$, so that $T_i(x_i)$ represents the execution time of running $x_i$ on $P_i$. In other words, the execution time of any distribution like $X$ is equal to the execution time of one the problem sizes in $X$ (i.e. $x_i \in X$) which has maximum execution time. Since we have $p$ time functions with a cardinality of $m$, there exist up to $m \times p$ data points with different execution times. Therefore, one can conclude that the number of solutions with different execution times cannot go beyond $m \times p$.

On the other hand, regarding the definition of Pareto-optimality, we know that values for energy and performance are unique in any Pareto-optimal set where there is not two solutions in one set where either their dynamic energy consumptions or their execution times are the same. Since there exist up to $m \times p$ distinct execution time, the cardinality of the Pareto-optimal set cannot exceed $m \times p$.
\textit{End of Proof}.

\begin{lemma}	\label{lemma:mergeComplexity}
		The computational complexity of the function \Call{MergePartialParetoes}{} for building Pareto-optimal solutions of a given node in a solution tree for a heterogeneous platform including $p$ discrete dynamic energy and $p$ performance functions with a cardinality of $m$ is equal to $O(m^2 \times p \times \log_2(m \times p))$.   
\end{lemma}

\textit{Proof.} Consider a given node $N$ at a level $L_c$ of a solution tree. As explained in the main manuscript, the node has generally up to $m + 1$ children. Regarding Lemma \ref{lemma:paretoSols}, each child of $N$ at $L_{c+1}$ has up to $m \times (p-c-1)$ Pareto-optimal solutions. Consider $\Psi_N$, a data structure of the type \emph{map}, storing the Pareto-optimal solutions of the node $N$. We know that the cardinality of $\Psi_N$ does not exceed $m \times (p-c)$ Pareto-optimal solutions (Lemma \ref{lemma:paretoSols}). To find the Pareto-optimal solutions of the node $N$, there are totally $(m + 1) \times (m \times (p-c-1))$ merged solutions which should be examined one by one so that inserting a merged solution in $\Psi_N$ has a complexity of $\log_2(m \times (p-c))$. Henceforth, the cost of processing and inserting all merged solutions is $(m + 1) \times (m \times (p-c-1)) \times \log_2(m \times (p-c))$ $\approxeq$ $O(m^2 \times p \times \log_2(m \times p))$. There exist around $(m + 1) \times (m \times (p-c-1)) - m \times (p-c)$ non-Pareto-optimal solutions which are being eliminated from $\Psi_N$ during the processing of the merged solutions. The elimination cost is totally $O(m^2 \times p)$. Therefore, the computational cost of merging all solutions for a given node in a search tree is equal to $O(m^2 \times p \times \log_2(m \times p))$. 
\textit{End of Proof}


\begin{lemma}	\label{lemma_1}
	The computational complexity of \emph{HEPOPTA\_Kernel} is $O(m^3 \times p^3 \times \log_2(m \times p))$.
\end{lemma}

\textit{Proof.} \emph{HEPOPTA\_Kernel} is an enhanced recursive algorithm, and therefore, its computational complexity can be related in terms of the number of its recursions. We will formulate the number of recursions using a trivial sample tree. 

Let's consider a workload $n$ executing on $5$ heterogeneous processors ($p = 5$). Suppose there exist five discrete performance, $T_i(x)$, and five discrete dynamic energy functions, $E_i(x)$ with a cardinality of $2$ ($m = 2$) where $x = \{\Delta x, 2\Delta x\}$ and $ i \in \{0,1, \cdots,4\}$. It should be noted that \emph{HEPOPTA\_Kernel} is able to deal with any granularity for workload sizes and considering the fix granularity size $\Delta x \in \mathbb{N}$ does not make the proof less general. Without loss of generality and for the sake of simplicity, we assume that execution time and dynamic energy consumption increase when problem size increases.

\begin{figure*}
	\centering
	\begin{tikzpicture} [scale=0.6,sloped,grow=right]
	\tikzset{level 1/.style={level distance=4cm,sibling distance=0cm}}
	\tikzset{level 2/.style={level distance=3.5cm,sibling distance=0.5cm}}
	\tikzset{level 3/.style={level distance=3.5cm,sibling distance=0.5cm}}
	\tikzset{level 4/.style={level distance=3.5cm,sibling distance=0.5cm}}
	\tikzset{level 5/.style={level distance=3.5cm,sibling distance=0cm}}
		\Tree	[.n \edge node[anchor=south] {0};
					[.n \edge node[anchor=south] {0};
						[.n \edge node[anchor=south] {0};
							[.n \edge node[anchor=south] {0};
								[.n \edge node[anchor=south] {n};
									\node [text width=0.7cm,align=center]{\textbf{0}};
								]
								\edge node[anchor=south] {$\Delta x$};
								[.$n-\Delta x$ \edge node[anchor=south] {$n-\Delta x$};
									\node [text width=0.7cm,align=center]{\textbf{0}};
								]
								\edge node[anchor=south] {$2\Delta x$};
								[.$n-2\Delta x$ \edge node[anchor=south] {$n-2\Delta x$};
									\node [text width=0.7cm,align=center]{\textbf{0}};
								]
							]
							\edge node[anchor=north] {$\Delta x$};
							[.$n-\Delta x$ \edge node[anchor=south] {0};
								[.$n-\Delta x$ \edge node[anchor=south] {$n-\Delta x$};
									\node [text width=0.7cm,align=center]{\textbf{0}};
								]
								\edge node[anchor=south] {$\Delta x$};
								[.$n-2\Delta x$ \edge node[anchor=south] {$n-2\Delta x$};
									\node [text width=0.7cm,align=center]{\textbf{0}};
								]
								\edge node[anchor=south] {$2\Delta x$};
								[.$n-3\Delta x$ \edge node[anchor=south] {$n-3\Delta x$};
									\node [text width=0.7cm,align=center]{\textbf{0}};
								]
							]
							\edge node[anchor=south] {$2\Delta x$};
							[.$n-2\Delta x$ \edge node[anchor=south] {0};
								[.$n-2\Delta x$ \edge node[anchor=south] {$n-2\Delta x$};
									\node [text width=0.7cm,align=center]{\textbf{0}};
								]
								\edge node[anchor=south] {$\Delta x$};
								[.$n-3\Delta x$ \edge node[anchor=south] {$n-3\Delta x$};
									\node [text width=0.7cm,align=center]{\textbf{0}};
								]
								\edge node[anchor=south] {$2\Delta x$};
								[.$n-4\Delta x$ \edge node[anchor=south] {$n-4\Delta x$};
									\node [text width=0.7cm,align=center]{\textbf{0}};
								]
							]
						]
						\edge node[anchor=south] {$\Delta x$};
						[.$n-\Delta x$ \edge node[anchor=south] {0};
							[.\node [red]{$n-\Delta x$};
							]
							\edge node[anchor=south] {$\Delta x$};
							[.\node [red]{$n-2\Delta x$};
							]
							\edge node[anchor=south] {$2\Delta x$};
							[.$n-3\Delta x$ \edge node[anchor=south] {0};
								[.$n-3\Delta x$ \edge node[anchor=south] {$n-3\Delta x$};
									\node [text width=0.7cm,align=center]{\textbf{0}};
								]
								\edge node[anchor=south] {$\Delta x$};
								[.$n-4\Delta x$ \edge node[anchor=south] {$n-4\Delta x$};
									\node [text width=0.7cm,align=center]{\textbf{0}};
								]
								\edge node[anchor=south] {$2\Delta x$};
								[.$n-5\Delta x$ \edge node[anchor=south] {$n-5\Delta x$};
									\node [text width=0.7cm,align=center]{\textbf{0}};
								]
							]
						]
						\edge node[anchor=south] {$2\Delta x$};
						[.$n-2\Delta x$ \edge node[anchor=south] {0};
							[.\node [red]{$n-2\Delta x$};
							]
							\edge node[anchor=south] {$\Delta x$};
							[.\node [red]{$n-3\Delta x$};
							]
							\edge node[anchor=south] {$2\Delta x$};
							[.$n-4\Delta x$ \edge node[anchor=south] {0};
								[.$n-4\Delta x$ \edge node[anchor=south] {$n-4\Delta x$};
									\node [text width=0.7cm,align=center]{\textbf{0}};
								]
								\edge node[anchor=south] {$\Delta x$};
								[.$n-5\Delta x$ \edge node[anchor=south] {$n-5\Delta x$};
									\node [text width=0.7cm,align=center]{\textbf{0}};
								]
								\edge node[anchor=south] {$2\Delta x$};
								[.$n-6\Delta x$ \edge node[anchor=south] {$n-6\Delta x$};
									\node [text width=0.7cm,align=center]{\textbf{0}};
								]
							]
						]
					]
					\edge node[anchor=south] {$\Delta x$};
					[.$n-\Delta x$ \edge node[anchor=south] {0};
						[.\node [red]{$n-\Delta x$};
						]
						\edge node[anchor=south] {$\Delta x$};
						[.\node [red]{$n-2\Delta x$};
						]
						\edge node[anchor=south] {$2\Delta x$};
						[.$n-3\Delta x$ \edge node[anchor=south] {0};
							[.\node [red]{$n-3\Delta x$};
							]
							\edge node[anchor=south] {$\Delta x$};
							[.\node [red]{$n-4\Delta x$};
							]
							\edge node[anchor=south] {$2\Delta x$};
							[.$n-5\Delta x$ \edge node[anchor=south] {0};
								[.$n-5\Delta x$ \edge node[anchor=south] {$n-5\Delta x$};
									\node [text width=0.7cm,align=center]{\textbf{0}};
								]
								\edge node[anchor=south] {$\Delta x$};
								[.$n-6\Delta x$ \edge node[anchor=south] {$n-6\Delta x$};
									\node [text width=0.7cm,align=center]{\textbf{0}};
								]
								\edge node[anchor=south] {$2\Delta x$};
								[.$n-7\Delta x$ \edge node[anchor=south] {$n-7\Delta x$};
									\node [text width=0.7cm,align=center]{\textbf{0}};
								]
							]
						]
					]
					\edge node[anchor=south] {$2\Delta x$};
					[.$n-2\Delta x$ \edge node[anchor=south] {0};
						[.\node [red]{$n-2\Delta x$};
						]
						\edge node[anchor=south] {$\Delta x$};
						[.\node [red]{$n-3\Delta x$};
						]
						\edge node[anchor=south] {$2\Delta x$};
						[.$n-4\Delta x$ \edge node[anchor=south] {0};
							[.\node [red]{$n-4\Delta x$};
							]
							\edge node[anchor=south] {$\Delta x$};
							[.\node [red]{$n-5\Delta x$};
							]
							\edge node[anchor=south] {$2\Delta x$};
							[.$n-6\Delta x$ \edge node[anchor=south] {0};
								[.$n-6\Delta x$ \edge node[anchor=south] {$n-6\Delta x$};
									\node [text width=0.7cm,align=center]{\textbf{0}};
								]
								\edge node[anchor=south] {$\Delta x$};
								[.$n-7\Delta x$ \edge node[anchor=south] {$n-7\Delta x$};
									\node [text width=0.7cm,align=center]{\textbf{0}};
								]
								\edge node[anchor=south] {$2\Delta x$};
								[.$n-8\Delta x$ \edge node[anchor=south] {$n-8\Delta x$};
									\node [text width=0.7cm,align=center]{\textbf{0}};
								]
							]
						]
					]	
				]		
	\end{tikzpicture}
\caption{The \emph{HEPOPTA} solution tree for executing a sample set of five profiles ($p=5$), each contains 2 data points. The memorization technique is only considered to reduce the full search space of solutions.}
	\label{fig:search_tree_time_com}
\end{figure*}
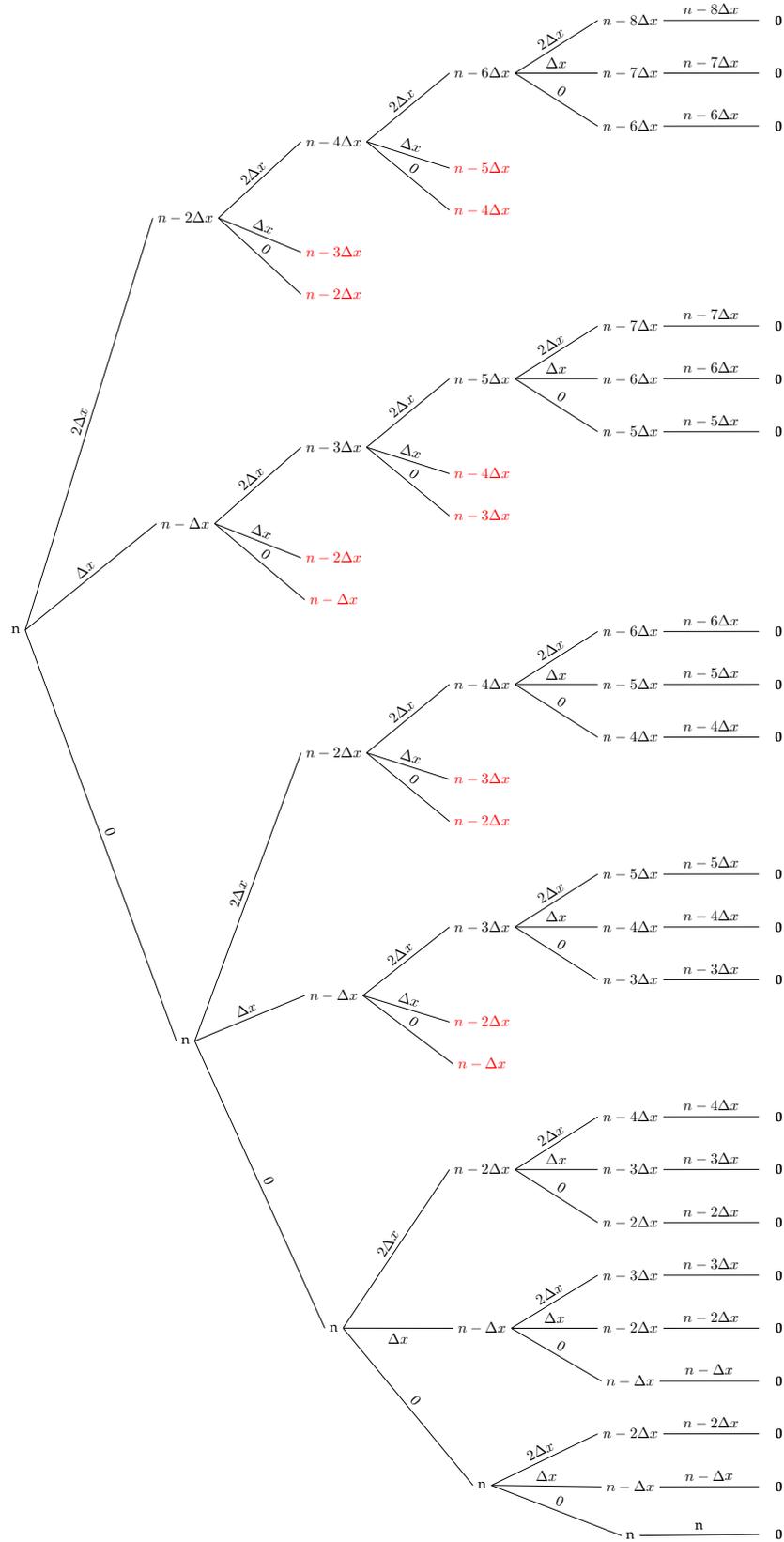

Figure \ref{fig:search_tree_time_com} shows the solution tree for finding the Pareto-optimal solutions for the workload $n$ on the five processors. Let's $n$ be greater than $8\Delta x$, the maximum possible workload which is subtracted from $n$ in this example. In the figure, red nodes represent ones have been already expanded in the same level, and their solutions are retrieved from $PMem$. For the sake of simplicity, the operation \emph{Cut} has not been employed.

According to the sample tree, the number of recursions (the number of nodes that their solutions do not retrieve form memory) in each level of solutions tree explored can be obtained using the Eq. \ref{eq:hepopta_recCalls}.

\begin{equation}	\label{eq:hepopta_recCalls}
    C\#(L) = \begin{cases}
    L \times m+1	  				& 0 \leq L < p-1\\
    C\#(p-2)\times (m + 1)	& L = p - 1\\
    \end{cases}
\end{equation}
where $L$ represents the level number.

The expanded form of Eq. \ref{eq:hepopta_recCalls} is shown in Eq. \ref{eq:hepopta_recCalls_simple}.

\begin{equation}	\label{eq:hepopta_recCalls_simple}
    C\#(L) = \begin{cases}
               L \times m+1	  										& 0 \leq L < p-1\\
               m^2 \times p - 2 \times m^2 + m \times p - m + 1		& L = p - 1\\
           \end{cases}
\end{equation}

That is, the total number of recursive calls is equal to $\sum_{L=0}^{p-1}(C\#(L))$ which is equal to $O(m \times p ^ 2 + m ^ 2 \times p)$.

In addition, the number of nodes in each level that their results are retrieved from $PMem$ is formulated in Eq. \ref{eq:hepopta_mem_total}.

\begin{equation}	\label{eq:hepopta_mem_total}
	\begin{split}
	&\text{Memory\#(L)} = (C\#(L-1) - 1) \times m \\ 
	&\qquad \qquad \quad~~= (m^2) \times (L - 1), \quad 1 \leq L \leq p - 2\\
	\end{split}
\end{equation}

Since $PMem$ saves the solutions which are found on levels $1$ to $p-2$, the total number of nodes that their solutions are saved (nodes in red in the figure) is equal to $\sum_{L=1}^{p-2} Memory\#(L) = O(m^2 \times p^2)$. The complexity of \Call{ReadParetoMem}{} is $O(1)$. Therefore, the computational cost for retrieving all solutions from $PMem$ is equal to $O(m^2 \times p^2)$.

The function \Call{MergePartialParetoes}{} is invoked after exploring all children of any node (nodes in black in Figure \ref{fig:search_tree_time_com}) in levels $\{L_0,\cdots,L_{p-2}\}$. Regarding Lemma \ref{lemma:mergeComplexity} and Eq. \ref{eq:hepopta_recCalls}, the total cost of all \Call{MergePartialParetoes}{} calls is equal to $\sum_{L=0}^{p-2} (L \times m + 1) \times (m^2 \times p \times \log_2(m \times p)) = O(m^3 \times p^3 \times \log_2(m \times p))$.


The computational complexity of \emph{HEPOPTA\_Kernel} can be summarized as follows:

\begin{equation*}
\begin{split}
\text{Complexity(\emph{HEPOPTA\_Kernel})} = &\text{O(recursive calls of \emph{HEPOPTA\_Kernel})} + \\
& \text{O($PMem$ solutions)} + \\
& \text{O(\Call{MergePartialParetoes}{} calls)}.
\end{split}
\end{equation*}
which equals:
\begin{equation*}
\begin{split}
\text{Complexity(\emph{HEPOPTA\_Kernel})} = & O(m \times p^2 + m^2 \times p)+ \\
& O(m^2 \times p^2) +\\
& O(m^3 \times p^3 \times \log_2(m \times p)) \\
& = O(m^3 \times p^3 \times \log_2(m \times p)).
\end{split}
\end{equation*}

\begin{proposition}
	The computational complexity of \emph{HEPOPTA} is $O(m^3 \times p^3 \times \log_2(m \times p))$.
\end{proposition}

\textit{Proof.} \emph{HEPOPTA} consists of following main steps:
\begin{itemize}
	\item \textbf{Sorting:} There exist $p$ discrete performance and $p$ discrete dynamic energy profiles with a cardinality of $m$. The complexity to sort all of them is $O(p \times m \times \log_2 m)$.
	\item \textbf{Initializing energy threshold $\varepsilon$:} Obtaining energy threshold involves two steps: (i) invoking \emph{HPOPTA} with a complexity of $O (m^3 \times p^3)$ followed by (ii) calculating the energy threshold $\varepsilon$ with a complexity of $O (p)$. Therefore, the complexity of this step is equal to $O (m^3 \times p^3)$.
	\item \textbf{Finding size thresholds:} To find the size threshold a given level $L_i$, $i \in [0,p-1]$, all data points, existing in $e_i(x)$ with dynamic energy consumptions greater than $\varepsilon$ should be examined in a complexity of $O(m)$. Therefore, finding $p$ size thresholds has a complexity of $O(p \times m)$.
	\item \textbf{Memory initialization:} In this step, all $(n+1) \times (p-2)$ cells of $PMem$ are initialized with a complexity of $O(n \times p)$.
	\item \textbf{Kernel invocation:} According to Lemma \ref{lemma_1}, the complexity of \emph{HEPOPTA\_Kernel} is $O(m^3 \times p^3 \times \log_2(m \times p))$.
\end{itemize}

Thus, the computational complexity of \emph{HEPOPTA} is equal to the summation of all these steps, which is equal to $O(m^3 \times p^3 \times \log_2(m \times p))$.
\textit{End of Proof}.

\begin{proposition}	\label{prop_mem_use}
	The total memory consumption of \emph{HEPOPTA} is $O(n \times m \times p^2)$.
\end{proposition}

\textit{Proof.} 
\emph{HEPOPTA} uses memory to store following information:
\begin{itemize}
	\item \textbf{energy functions:} There are $p$ discrete energy functions with cardinality of $m$. We store both size-sorted (sorted by problem size) and energy-sorted (sorted by the amount of dynamic energy consumption) functions. These function are stored in $2 \times p \times m$.
	\item \textbf{time functions:} There are $p$ discrete time functions with cardinality of $m$. We store both size-sorted (sorted by problem size) and time-sorted (sorted by execution time) functions. These function are stored in $2 \times p \times m$.
	\item \textbf{$\Psi_{EP}$:} Regarding Lemma \ref{lemma:paretoSols}, the maximum number of Pareto-optimal solutions are $m \times p$. Since the workload distribution of each solution, including $p$ elements, is stored in $\Psi_{EP}$, the maximum size of the set $\Psi_{EP}$ is $O (m \times p^2)$.
	\item \textbf{PMem:} This is a matrix consisting of $(p-2) \times (n+1)$ cells. Each cell stores up to $m \times p$ Pareto-optimal solutions (Lemma \ref{lemma:paretoSols}). Therefore, the memory usage of $PMem$ is equal to $O(n \times m \times p^2)$.
	\item \textbf{Memory consumption of \emph{HPOPTA}:} The memory usage of \emph{HPOPTA} is $O (p \times (m + n))$.
	\item \textbf{$X_{cur}$}: This is an array of $p$ elements to store the problem sizes assigned to each processor by the current solution.
	\item \textbf{partsVec}: This is a vector of size $O (m)$ storing the problem sizes given to a processor where results in a solution. There exists $p-1$ partsVecs, one per level. That it total consumed memory is equal to $O(m \times p)$.
\end{itemize}

Thus, an upper bound for the total memory usage of \emph{HEPOPTA} is equal to $O(n \times m \times p^2)$. 
\textit{End of Proof}.

\subsection{Definition of Pareto-optimal Solutions for Dynamic Energy and Execution Time}

Suppose there exists a given workload $n$ executing on $p$ processors using a workload distribution $X^* = \{x_0^*,x_1^*,\cdots,x_{p-1}^*\}$ where $\sum_{i=0}^{p-1}x_i^* = n$, $E_D(X^*) = \sum_{i=0}^{p-1} e_i(x_i^*)$ is the dynamic energy consumption of the distribution, and $T_E(X^*) = \max_{i=0}^{p-1} t_i(x_i^*)$ represents its execution time.

Suppose $S$ represents the feasible set of distributions. According to the definition of Pareto-optimality, the distribution $X^*$ would be a Pareto-optimal solution if its dynamic energy consumption ($E_D(X^*)$) and execution time ($T_E(X^*)$) satisfy Eq. \ref{eq:pareto-def}. In this equation, $X = \{x_0, x_1, \cdots, x_{p-1}\} \in S$ represents any workload distribution for $n$.

\begin{equation}	\label{eq:pareto-def}
	\begin{split}
		& \nexists X \in S~|~E_D(X) \le E_D(X^*) \land T_E(X) < T_E(X^*) \\
		& AND \\
		& \nexists X \in S~|~T_E(X) \le T_E(X^*) \land E_D(X) < E_D(X^*)
	\end{split}
\end{equation}

Equation \ref{eq:pareto-def} means that there does not exist any objective vector $(E_D(X),T_E(X))$ for which all the objective vector values are less than Pareto-optimal vector $(E_D(X^*),T_E(x^*))$. In fact, there is no other solution which dominates $X^*$.

\begin{lemma}	\label{lemma:1}
	For each non-Pareto-optimal workload distribution $X$ with the objective vector $(E_D(X),T_E(X))$, there is at least one Pareto-optimal solution $X^*$ where either $E_D(X^*) \le E_D(X)$ and $T_E(X^*) < T_E(X)$ or $T_E(X^*) \le T_E(X)$ and $E_D(X^*) < E_D(X)$.
\end{lemma}

\textit{Proof.} Regarding the Eq. \ref{eq:pareto-def}, for each non-Pareto-optimal solution $X$ exists at least one solution $Y$ in the objective space where $E_D(Y) \le E_D(X) \land T_E(Y) < T_E(X)$ or $T_E(Y) \le T_E(X) \land E_D(Y) < E_D(X)$. The point $X$ is called \emph{dominant point}. If the dominant point $Y$ is a Pareto-optimal solution, the correctness of the lemma is proven. But if not so, there exists another dominant point so that dominates $Y$. The process of finding dominant points can be recursively repeated. The recursion will eventually terminate because the two objectives execution time and dynamic energy consumption are \textit{finite positive} parameters and their values gradually decrease during this recursive process. That is, the recursive process finally reaches a given solution $X^*$ which cannot be dominated by any other solution. According to the definition of Pareto-optimal solutions, the solution $X^*$ should be a member of the Pareto-optimal set for execution time and dynamic energy ($\Psi_{EP}$). Therefore, the correctness of the lemma \ref{lemma:1} is proven.
\textit{End of Proof}.

\subsection{Pareto-front Solutions for Total Energy and Execution Time}

In this section, we will prove how to build Pareto-optimal solutions for execution time and total energy ($\Psi_{TP}$) by using Pareto-optimal solutions for execution time and dynamic energy ($\Psi_{EP}$).

\begin{proposition} \label{prop:1}
	There is no workload distribution $X$ such that $X \notin \Psi_{EP}$ but $X \in \Psi_{TP}$.
\end{proposition} 

\textit{Proof.} Referring to Lemma \ref{lemma:1}, if a solution $X$ is not in $\Psi_{EP}$ then there exists a Pareto-optimal solution $X^*$ such that either $E_D(X^*) \le E_D(X)$ and $T_E(X^*) < T_E(X)$ or $T_E(X^*) \le T_E(X)$ and $E_D(X^*) < E_D(X)$. Since total energy is a function of execution time and dynamic energy, it can be deducted that $E_T(X^*) < E_T(X)$. Since $T_E(X^*) < T_E(X)$ and either $E_T(X^*) < E_T(X)$ or $E_T(X^*) \le E_T(X)$, the objective vector $(E_T(X^*), T_E(X^*))$ dominates $(E_T(X), T_E(X))$. Therefore, it can be concluded that feasible solutions which are not a remember of $\Psi_{EP}$ cannot be a member of $\Psi_{TP}$.
\textit{End of Proof}.

\begin{proposition} \label{prop:2}
	If $X_{opt}$ is a workload distribution minimising total energy consumption, its corresponding objective vector, $(E_D(X_{opt}),T_E(X_{opt}))$, is necessarily a member of Pareto-optimal set for dynamic energy and performance.
\end{proposition}

\textit{Proof.} We categorize all points in the feasible set of distributions into two distinct groups: (i) solutions existing in the Pareto-optimal set $\Psi_{EP}$, and (ii) solutions are not Pareto-optimal. Regarding Lemma \ref{lemma:1}, for each non-Pareto-optimal solution $X$, there is at least one solution $X^*$ in Pareto-optimal set dominating $X$. Since total energy is a function of execution time and dynamic energy, it can be concluded that 

$$\forall x \notin~\Psi_{EP}, \exists X^* \in~\Psi_{EP}$$
$$where$$
$$E_T(X^*) < E_T(X)$$

That is, the solution which minimizes total energy must be a member of Pareto-optimal set.
\textit{End of Proof}.

\subsection{Complexity of \emph{HTPOPTA}}

\begin{proposition}
	The computational complexity of \emph{HTPOPTA} is $O(m^3 \times p^3 \times \log_2(m \times p))$.
\end{proposition}

\textit{Proof.} To find globally Pareto-optimal solutions for total energy and performance, \emph{HTPOPTA}, first, invokes \emph{HEPOPTA} for obtaining $\Psi_{EP}$, with a complexity of $O(m^3 \times p^3 \times \log_2(m \times p))$. As explained earlier, the number of solutions in $\Psi_{EP}$ does not exceed $m \times p$.

\emph{HTPOPTA} then calculates the total energy consumption of each solution in $\Psi_{EP}$ (up to $m \times p$ number of solutions). It inserts the new solutions into $\Psi_{TP}$ or updates existing ones. $\Psi_{TP}$ is defined as a data structure of the type \emph{map} to store Pareto-optimal solutions for total energy and performance. In the case of existing two solutions with equal total energy consumption and execution times, the solution with less active processors (processors with non-zero workloads) is chosen by \emph{HTPOPTA}. Inserting a solution in $\Psi_{TP}$ has a logarithmic computational complexity, and determining solutions with less active processors has a complexity of $O(p)$. Therefore, the cost of inserting and updating up to $m \times p$ solutions in $\Psi_{TP}$ is $(m \times p) \times (\log_2(m \times p) + p)$ $\approxeq$ $O(m \times p^2)$. In addition, the computational complexity for eliminating all non-Pareto-optimal solutions from $\Psi_{TP}$ is $O(m \times p)$.

Therefore, the total computational cost to calculate $\Psi_{TP}$ is equal to $O(m^3 \times p^3 \times \log_2(m \times p))$.
\textit{End of Proof}.

\begin{proposition}
	The total memory consumption of \emph{HTPOPTA} is $O(n \times m \times p^2)$.
\end{proposition}

\textit{Proof.} 
\emph{HTPOPTA} uses memory to store following information:
\begin{itemize}
	\item \textbf{energy functions:} There are $p$ discrete energy functions with a cardinality of $m$. We store both size-sorted (sorted by problem size) and energy-sorted (sorted by the amount of dynamic energy consumption) functions. These function are stored in $2 \times p \times m$.
	\item \textbf{time functions:} There are $p$ discrete time functions with a cardinality of $m$. We store both size-sorted (sorted by problem size) and time-sorted (sorted by execution time) functions. These function are stored in $2 \times p \times m$.
	\item \textbf{Required memory by \emph{HEPOPTA:}} The total memory consumption of \emph{HEPOPTA} is $O(n \times m \times p^2)$.
	\item \textbf{$\Psi_{TP}$:} Regarding the Lemma \ref{lemma:paretoSols}, the maximum number of Pareto-optimal solutions is equal to $m \times p$. Since the workload distribution of each solution, involving $p$ elements, is stored in $\Psi_{TP}$, the maximum size of $\Psi_{TP}$ is $O (m \times p^2)$.
\end{itemize}

Thus, total memory usage of \emph{HTPOPTA} is equal to $O(n \times m \times p^2)$. 
\textit{End of Proof}.

\ifCLASSOPTIONcompsoc
  \section*{Acknowledgements}
\else
  \section*{Acknowledgement}
\fi

This publication has emanated from research conducted with the financial support of Science Foundation Ireland (SFI) under Grant Number 14/IA/2474.

\bibliographystyle{IEEEtran}
\bibliography{IEEEabrv,paper}

<<<<<<< HEAD
\vfill

\end{document}